%% file: main.tex
\documentclass[9pt]{sig-alternate-05-2015}

\setcopyright{acmcopyright}

\usepackage{listings}
\usepackage{tabularx}
\usepackage{booktabs}
\usepackage{multirow}
\usepackage{threeparttable}
\usepackage{caption}
\usepackage{subfig}

\usepackage{fancyhdr}

\usepackage[normalem]{ulem}

\usepackage{mathtools}

\newtheorem{theorem}{Theorem}

\newtheorem{definition}{Definition}

\usepackage[normalem]{ulem}
\usepackage{algorithm}
\usepackage{algpseudocode}

\usepackage{graphicx}

\usepackage{tikz}
\usepackage{verbatim}
\usetikzlibrary{arrows, shapes}

\usepackage{amssymb}

\lstdefinestyle{mystyle}{
	basicstyle=\footnotesize,
	breakatwhitespace=false,         
	keepspaces=true,                 
	showspaces=false,                
	showstringspaces=false,
	showtabs=false,                  
	tabsize=2
}

\begin{document}

\title{A Graph-based Model for GPU Caching Problems}

\numberofauthors{1}

\author{
	\alignauthor
	Lingda Li \hfil Ari B. Hayes \hfil Stephen A. Hackler \hfil Eddy Z. Zhang \\Mario Szegedy \hfil Shuaiwen Leon Song$^\star$\\
	\affaddr{Department of Computer Science, Rutgers University}\\
	\affaddr{$^\star$Pacific Northwest National Lab}\\
	\email{lingda.li@cs.rutgers.edu \hfil arihayes@cs.rutgers.edu \hfil s.hackler@rutgers.edu \\\{eddy.zhengzhang, szegedy\}@cs.rutgers.edu \hfil Shuaiwen.Song@pnnl.gov}
}

\maketitle





\lstset{style=mystyle}

\input{intro}

\input{probdef}

\input{tech}

\input{practical_new}

\input{eval_new}

\input{rel}

\section{Conclusion}
\label{sec:conclusion}


In this paper, we propose a task partition technique to improve data sharing among different GPU threads.
We use data-affinity graphs to model data sharing and map task scheduling problem to an edge partition problem.
This is the first time the \emph{edge-centric} model is used for GPU cache
performance modeling. We propose a novel partition algorithm based on the
\emph{edge-centric} model, and our algorithm provides high quality task schedule
and yet is low-overhead. 
We also provide rigorous proof for the analytical bound of our algorithm.
Our experiments show that our method can improve data sharing and thus performance significantly for various GPU applications.


\bibliographystyle{abbrv}
\bibliography{references}

\end{document}

%% file: intro.tex
\section{Introduction}
\label{sec:intro}

A GPU is a massively parallel computational accelerator that is equipped with
hundreds or thousands of cores. A single GPU can provide more than 4 Teraflops
single precision performance at its peak, however, the maximum memory throughput
of a GPU card is around 200 GB/s. Such a gap usually prevents GPU's computation power from being fully harnessed. 

A cache is a layer in between GPU's computation units and memory units, which
is a fast but (relatively) small storage for frequently accessed data. The
modern GPUs are equipped with cache to improve program performance
\cite{FermiWhitePaper, KeplerWhitePaper, MaxwellWhitePaper}. The
throughput of the first level cache on GPUs is close to the throughput of
computation unit and thus effective use of cache is critical to performance.
While regular applications can take advantage of cache by classical
transformations such as loop tiling and pipelining, instruction
scheduling and etc, it is not so straightforward how irregular
applications can best utilize cache.

\begin{figure}[h!]
	\centering
	\includegraphics[width=0.8\linewidth]{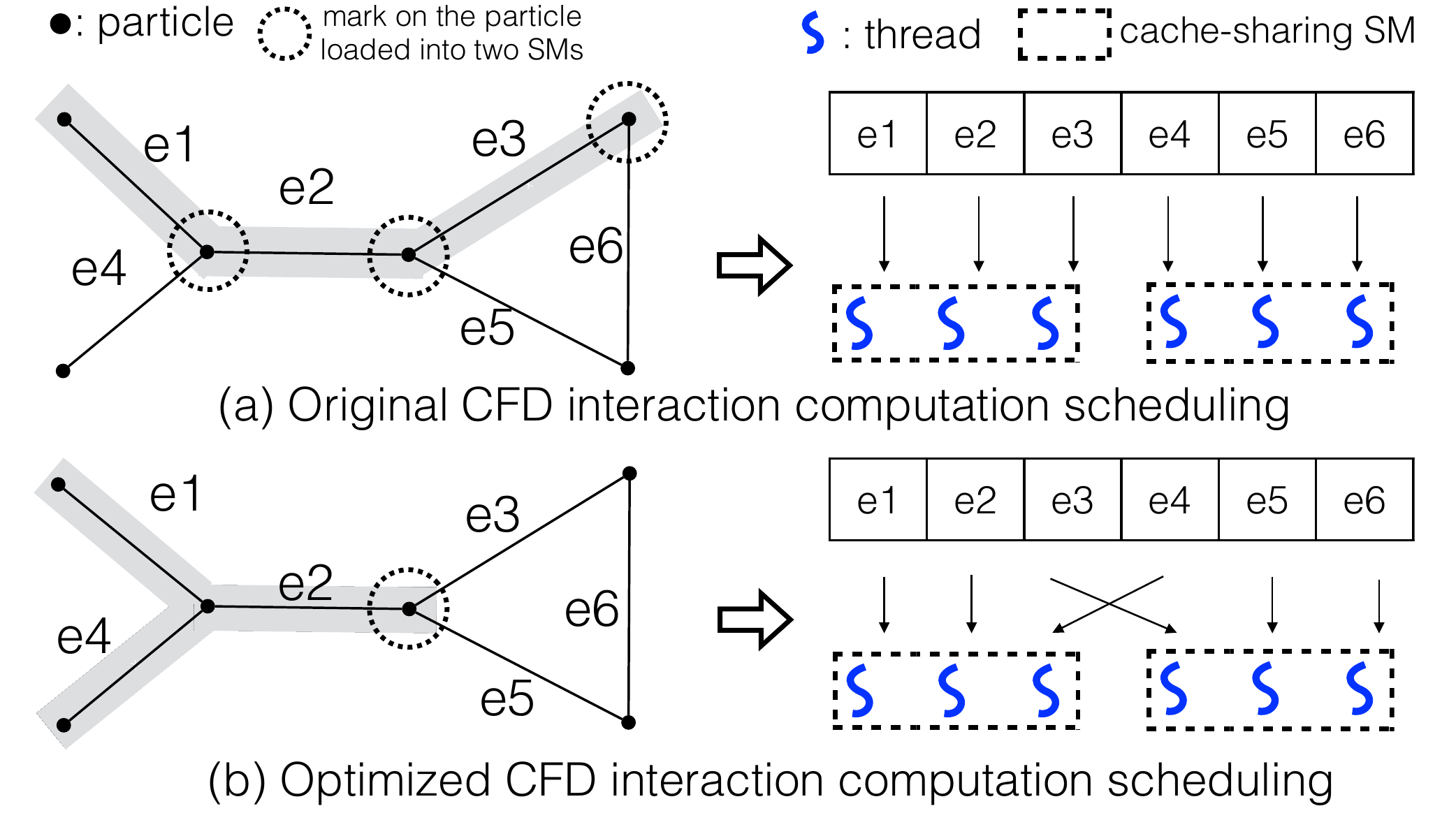}
	\caption{Mapping of \emph{cfd} interaction computation into two streaming multiprocessors. Assuming every SM has three threads. In (a) there are three particle needs to be loaded into two SMs while in (b) there is only one. }
	\label{fig:mot}
\end{figure}

In this paper, we look at irregular GPU applications and focus on improving
\emph{shared cache} performance. A GPU consists of a set of streaming multiple
processors (SMs or SMXs). Every SM has a fast shared cache. Threads that run on
a SM can maximize data sharing in cache and minimize communication to
off-chip memory.  We show an example of how threads can be effectively scheduled
to minimize off-chip memory communication in Figure \ref{fig:mot}.  

We use the computation fluid dynamics (\emph{CFD}) application
\cite{Corrigan+:AIAA09, Corrigan+:IJNMF11} as an example (Figure \ref{fig:mot}).
In \emph{cfd}, the main computation is to calculate interaction between two
adjacent particles and apply the interaction information to obtain the state of the
particles in the next time step.
Figure \ref{fig:mot} shows the interaction among six particles as a graph on the
left. Each node in the graph represents a particle. Each edge represents an
interaction, which is taken care of by one thread. Figure \ref{fig:mot}(a) shows
one way to schedule the threads, with interactions {\em e1,e2,e3} mapped to one
SM, requiring 9 loads from the memory (assuming one load is for every distinct
particle). In this case, three particles need to be fetched twice from memory. However, the thread schedule in figure
\ref{fig:mot}(b) indicates that, with {\em e1,e2,e4} mapped into one SM, only one
particle needs to be fetched from memory twice, and the particle that is shared
by {\em e1, e2, e4} only needs to be fetched into cache once. All
together the second case requires 7 loads from memory. 
According to our study, for all three inputs of \emph{CFD} from the Rodinia benchmark suite \cite{Che+:IISWC09}, an average of 73.4\% of all particle loads are redundant under the default task scheduling.
This example demonstrates that there is a great potential to improve data sharing through careful task scheduling.

The graph-based model is used and extensively studied in the past \cite{Hendrickson+:PC00, gonzalez2012powergraph,
malewicz2010pregel, low2012distributed} for irregular
application on parallel CPU processors, due to its efficiency in capturing the
relationship between data and computation. In this paper, we also used a
graph-based model to tackle the shared cache problem for irregular GPU applications.  

The GPU shared cache problem brings several new challenges which cannot be
handled well by previous graph-based methods.
First of all, the expected task partitioning algorithm must have low time overhead, due to the high computation throughput of GPU.
This constraint makes complex schemes like hypergraph partitioning \cite{Hendrickson+:PC00} infeasible.
Secondly, we expect high partition equality in terms of task schedule
(partitioning), since up to thousands of concurrent threads share only a few KB of cache space, making random or greedy partitioning methods (e.g., techniques in PowerGraph \cite{gonzalez2012powergraph}) impractical.
Thirdly, the task partition results should be highly balanced because of the GPU's
single instruction multiple thread (SIMT) execution model, in which every thread
executes the same code but on different input. The execution time of a parallel
program depends on the execution time of the longest task. Imbalanced task
schedule may cause significant slowdown. This is different from CPU processors,
which can typically tolerate a relatively higher degree of imbalance because of
limited amount of parallelism \cite{malewicz2010pregel}.

To address the above problems, we propose to use an \emph{edge-centric}
graph-based model and we propose a fast task partition
algorithm for the \emph{edge-centric} model. Our partition algorithms has low
overhead compared with \emph{vertex-centric} model and yet provides high quality
balanced partition results. 
Our contributions can be summarized as follows:  

\begin{itemize}
\item This is the first \emph{edge-centric} model for GPU cache performance. It
overcomes the problems of traditional \emph{vertex-centric} model and captures
communication volume accurately. 
\item We propose a task partition algorithm for the \emph{edge-centric} model.
Our algorithm gives the best approximation bound we know so far. 
\item Our task partition algorithm is fast, often of orders of magnitude faster
than previous \emph{hypergraph} model and other \emph{edge-centric} partition
algorithm. 
\item Our algorithm is robust. It provides steady performance improvement
regardless of the type of cache that is used: software cache or hardware cache,
and regardless of the thread block sizes (work group size).
\end{itemize}

The rest of the paper is organized as follows.
Section \ref{sec:background} describes the background of this paper.
Section \ref{sec:tech} introduces our task partition model and algorithm.
Section \ref{sec:appl} describes how to apply our model in GPU programs.
Section \ref{sec:eval} shows our experimental environment and evaluation results.
Section \ref{sec:rel} discusses the related work, and Section \ref{sec:conclusion} concludes the paper.

%% file: probdef.tex
\section{Background}
\label{sec:background}

\begin{figure}[t]
	\centering
	\includegraphics[width=0.4\textwidth]{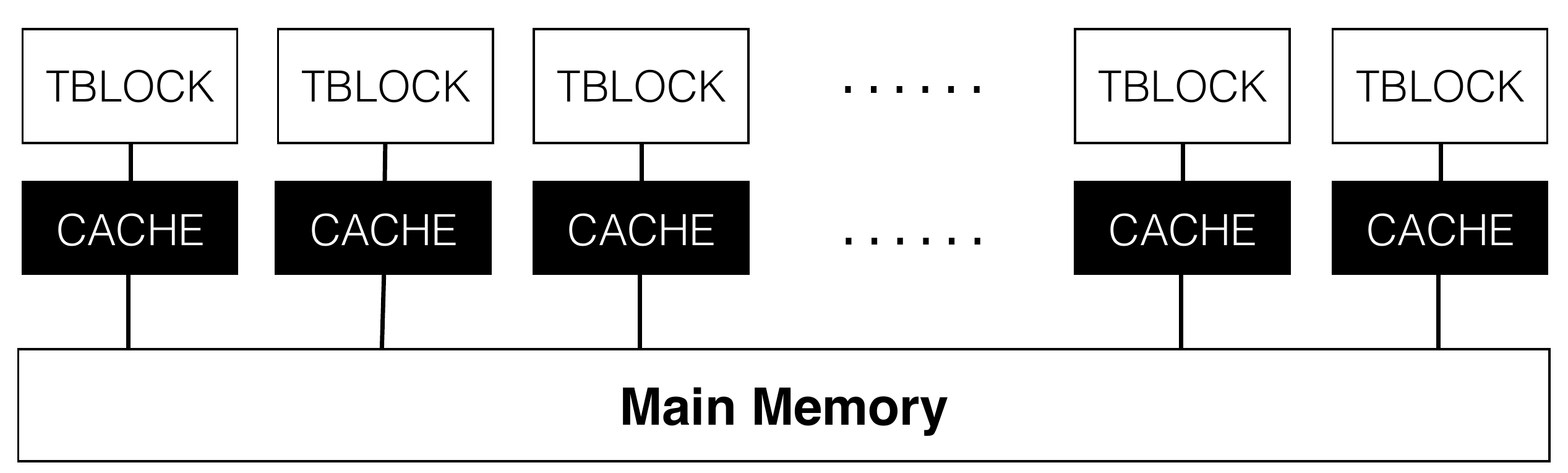}
	\caption{Abstract cache model for GPU cache architecture. TBLOCK represents a
thread block. }
	\label{fig:cachearch}
\end{figure}

We first describe the abstract memory model for a GPU architecture. As mentioned
earlier, a GPU is composed of a number of streaming multiprocessors (SMs or
SMXs). There is a cache on every SM, and threads running on the same SM share
this cache. For a GPU program, the minimal unit of threads that run on one SM is
called a thread block, also called \emph{cooperative thread arrays (CTAs)} using
NVIDIA CUDA terminology. The GPU threads are divided into blocks of the same
size. One thread block can run at most on one SM at one time.  Multiple thread
blocks can run simultaneously on one SM. A thread block
would acquire one SM and hold the SM until all its threads are finished. As one thread
block releases the SM, another thread block (if there is any) will acquire the
SM. As such, a thread block is also the minimal thread work group that share
cache. We show the abstract cache sharing model in Figure \ref{fig:cachearch}.
Every thread block uses a partition of the cache on one SM and would release it
when the thread block is finished. Therefore, there are as if as many virtual
cache as the number of thread blocks.

There are two types of caches on the SM: software cache and hardware cache.
Software cache is scratch-pad memory, which is also referred to as {\em shared memory} in NVIDIA CUDA terminology\footnote{We use NVIDIA CUDA terminology throughout the paper.}.
Software cache requires explicit load/store instructions to move frequently used data from/into
memory. Every thread block gets an exclusive and even partition of {\em shared
memory}. Similarly, the acquired partition of the software cache will not be released
until the owner thread block finishes. A thread block cannot see another thread
block's
software cache (shared memory) partition.

Hardware caches can be further classified into several categories based on their usage.
In recent GPU architectures, the L1 cache is only used to store local variables that are private to each thread \cite{KeplerWhitePaper, MaxwellWhitePaper}.
The texture cache can store shared data objects after the data objects are binded with texture memory.
Thus we use texture cache as the hardware cache to evaluate our task
schedule/partition algorithms. The hardware cache does not require software
to explicitly load frequently used data into cache, and has good
programmability. However, it may not always keep the right data in cache since
it does not have as much knowledge about the program as the programmer.
Therefore, there is the trade-off between programmability and performance. We
show the comparison between software cache and hardware cache in the evaluation
section.


Finally, there is a L2 cache which is shared by all SMs on a GPU.  A L2 cache
has a typical hit latency of hundreds of cycles (close to memory access latency
on GPU).
Because of its long latency and also because it is unified across all SMs like
the main memory, we
only focus on optimization of the first level cache on every SM. 





%% file: tech.tex
\section{Edge-Centric Model}
\label{sec:tech}

\subsection{Problem Definition}
\label{subsec:model}
We build a edge-centric graph model to describe the relationship between data objects and tasks.
In our model, we describe a data object as a vertex and a task as
an edge.

\begin{definition}
We define a data-affinity graph $D = (V, E)$ with the set of vertices $V$ and the set of
edges $E \subset V \times V$. Let $n$ and $m$ denote the number of vertices and the number of edges,
respectively. A vertex $v \in V$ represents a data object. An edge $e \in E$ with $e = (u,v)$
denotes a computation task that uses the two data objects $u$ and $v$.
\end{definition}

\begin{figure*}
\centering
\includegraphics[width=0.95\textwidth]{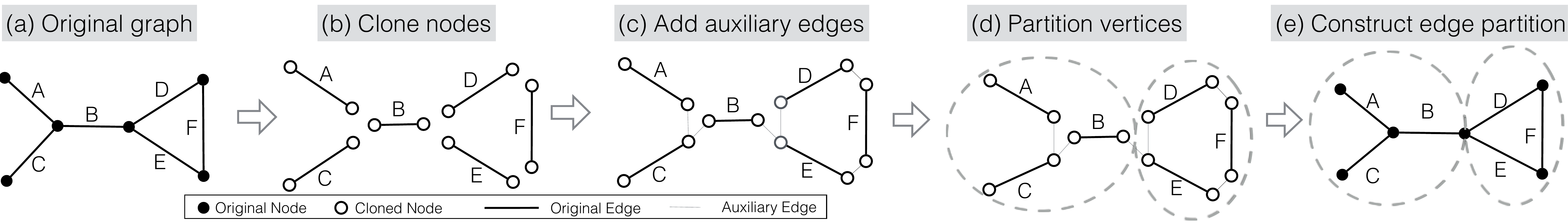}
\caption{Balanced edge partition problem converted to balanced vertex partition problem.}
\label{fig:epart}
\end{figure*}

Here every data object has the same attribute and size.
The definition of a data object typically depends on the semantics of the program.
One task could correspond to one or a group of operations in the program based on the semantics.
For instance, \emph{cfd} \cite{Che+:IISWC09} calculates the interaction of two
particles based on their density, energy, and momentum variation.
We group these three attributes of one particle into one data object. Thus one
task is defined as the calculation based on these two aggregate data objects
corresponding to this pair of particles.

Our goal is to partition all $m$ edges in the data-affinity graph evenly into $k$ clusters, such that every edge (task) is assigned to exactly one cluster (thread block).
In short, this is a balanced edge partitioning (EP) problem. The optimization objective is to maximize data reuse, which is equivalent to minimizing the number of times a vertex (data object) appears in different clusters (thread blocks).

\begin{definition}
\label{def:epproblem}
A vertex $v$ is cut if its incident edges belong to more than one cluster. We
define $c_v$ as the cost that $v$ is cut, which is the number of unique clusters
that $v$'s incident edges belong to $p_v$ subtracted by 1, $c_v = p_v - 1$. We define
a total vertex-cut cost $C$ which has to be minimized under the condition of edge balancing.
Let $C = \sum_{v \in V } (p_v - 1)$ and let $L_i$ denote the number of edges in
cluster i, the {\em k$-$way balanced edge partition (EP)} problem is defined as:

\begin{equation}\label{commcost}
\begin{array}{rlll}
\displaystyle \min & { C(x) = \sum_{v \in V} (p_v - 1)} \\
\textrm{s.t.} & \forall i \in [1...k] ~~ L_i(x)  = \frac{m}{k}  \\
							& x~is~a~valid~edge~partitioning
\end{array}
\end{equation}
\end{definition}

We use the same \emph{cfd} example from Section \ref{sec:intro} to illustrate the
edge partition model. Figure \ref{fig:epart}(a) shows an example of the data affinity graph for \emph{cfd}.
In this example, there are 6 interactions to be computed.
Assume a two-way balanced edge partition, i.e., $k=2$.
The total vertex cut cost of the partition in Figure \ref{fig:epart}(e) is one, since one vertex (data object) appears in two edge clusters.

The number of times the data objects appear in different edge partitions represents how often a data object is loaded into different thread blocks.
In the ideal case, each data object appears in only one partition.
In such scenarios, threads sharing data all reside in one thread block, and thus there is no redundant data access.
Each time one data object is shared by another thread block, one extra redundant data access is required.
Therefore, the total vertex cut $C(x) = \sum_{v \in V} (p_v-1)$ represents the number of redundant data accesses.
We also refer to it as the \emph{data reuse cost}, which we use to evaluate the quality of partitioning.

\subsection{Problem Transformation}
\label{subsec:alg}

It can be easily proved that the problem of balanced edge partitioning in
Definition \ref{def:epproblem} is NP-complete by reduction from existing NP-hard partition problems. For lack of space, we omit the proof here.

The edge partition problem is a non-traditional graph partition problem, and it lacks sophisticated solutions.
On the other hand, there is a rich exploration on both a theoretical foundation and practical solutions for the vertex partition problem.
We will show that we can convert the proposed balanced edge partition problem into the balanced vertex partition problem to leverage sophisticated vertex partition algorithms for efficient edge partitioning.
As far as we know, our work is the first to map a balanced edge partition problem into a balanced vertex partition problem.
We will prove that, not only does our solution work fast in practice, but also guarantees a polynomial algorithm with an worst case approximation factor of $(d_{max}-1)\mathcal{O}(\sqrt{\log{}m\log{}k})$, which is the best bound we know by far\footnote{$d_{max}$ is the maximum vertex degree in the data-affinity graph.}.

To convert the edge partition problem to a vertex partition problem, we first transform the data-affinity graph.
We replace every vertex in the graph 
with a set of new \emph{cloned vertices} such that every cloned vertex is connected to
one unique incident edge of the original vertex. The cloning process is shown in
Figure \ref{fig:epart}(b). Next, assuming the degree of the original vertex is $d$, we add $d-1$
\emph{auxiliary edges} to connect the set of $d$ cloned vertices to form a path.
This process of adding auxiliary edges is shown as an example in Figure \ref{fig:epart}(c).
Formally,

\begin{figure*}[htp]
\includegraphics[width=0.24\textwidth]{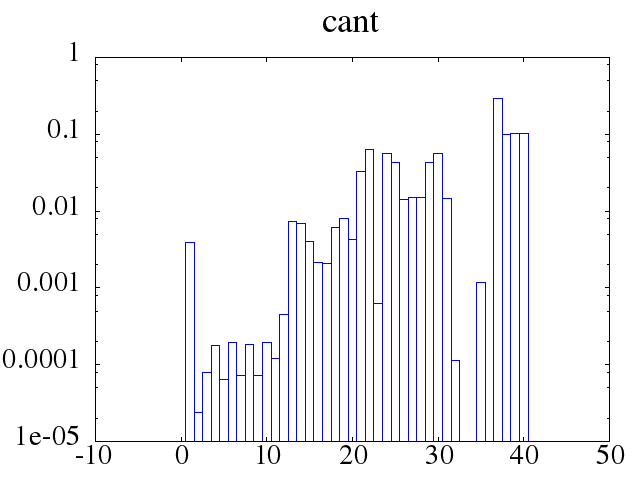}
\includegraphics[width=0.24\textwidth]{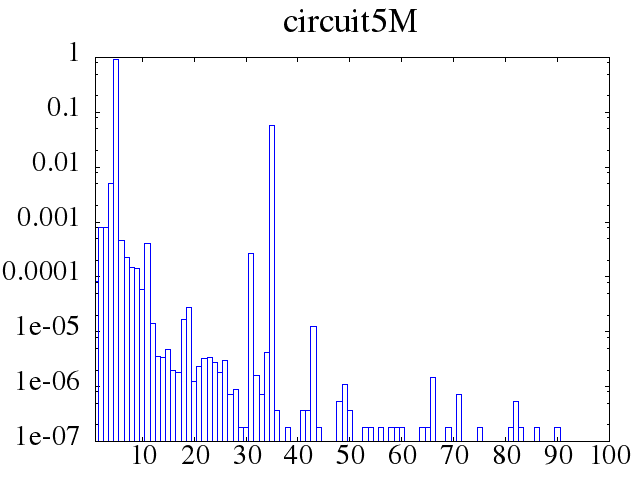}
\includegraphics[width=0.24\textwidth]{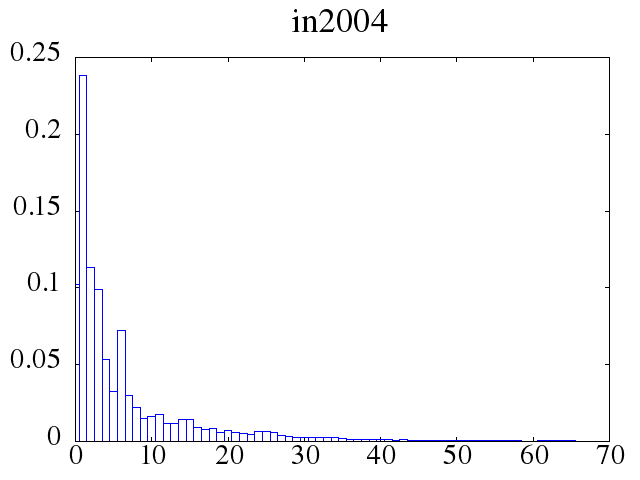}
\includegraphics[width=0.24\textwidth]{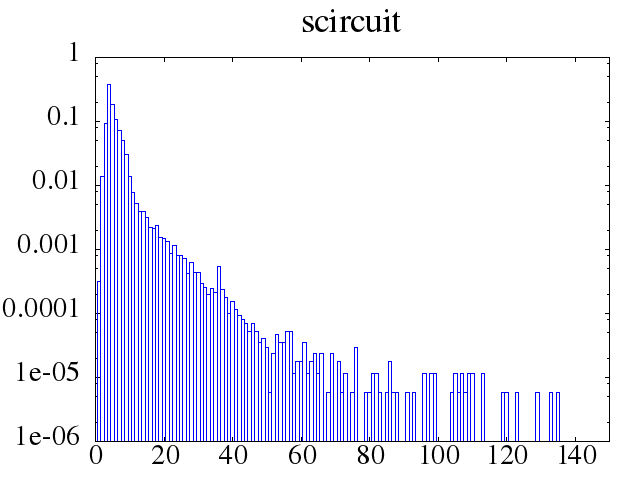}
\caption{Degree distribution: the frequency of a degree in the graphs for the
graphs \emph{cant}, \emph{circuit5M}, \emph{in2004}, \emph{scircuit}. For
readability, we only show part of the x range for circuit5M, in2004 and
scircuit. }
\label{fig:4pdf}
\end{figure*}

\begin{figure}[htp]
\includegraphics[width=0.23\textwidth]{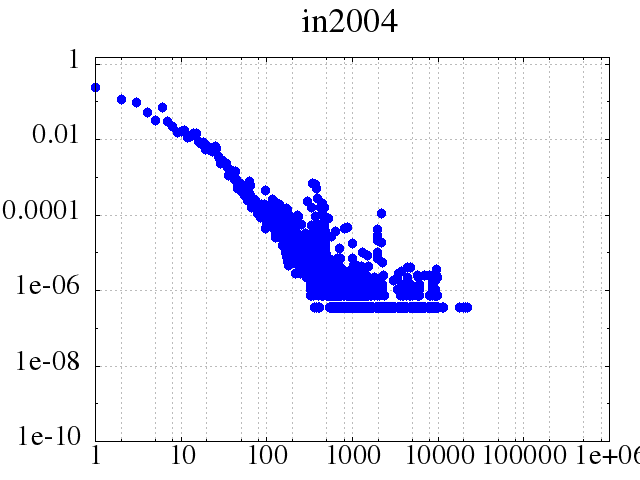}
\includegraphics[width=0.23\textwidth]{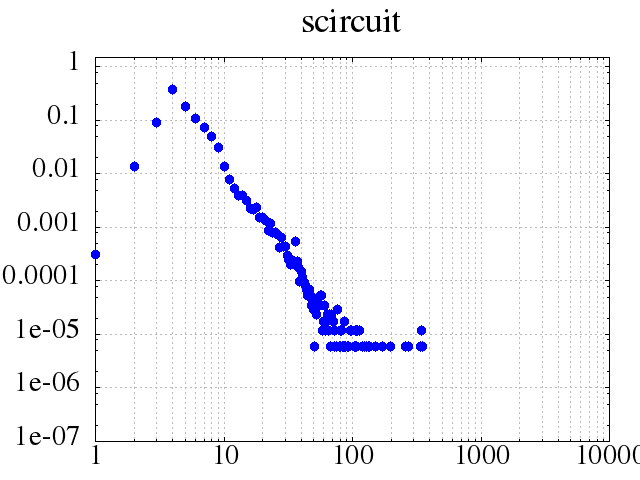}
\caption{Log-log scale of degree distribution for in2004 and scircuit}
\label{fig:loglog2}
\end{figure}

\begin{definition}
\label{def:graphconv}
We define a graph transformation as a clone-and-connect transformation if it
satisfies following conditions. Assume $t$ is a legal clone-and-connect
transformation and $D \xmapsto{t} D'$, where $D = (V, E)$ and  $D' = (V',E')$.
The transformed graph $D'$ needs to maintain the following characteristics. For every
vertex $v \in V$ of degree $d$, there are $d$ corresponding cloned vertices $v'_{1},\ldots v'_{d} \in
V'$. For every edge $e \in E$ with endpoints $u, v \in V$, there is a corresponding edge $e' \in
E'$ with endpoints $u'_i, v'_j \in V'$ such that $u'_i$ is the i-th cloned vertices of $u$, $v'_j$ the
j-th cloned vertex of $v$, and no cloned vertex is connected to more than one
original edge.
For every original vertex $v$ of degree $d$, its cloned set of vertices $v_i$, $i=1...d$ in $D'$ are connected into form a path using $d-1$ auxiliary edges, where a path is defined as a tree that contains only vertices of degree 2 and 1.
\end{definition}

By Definition \ref{def:graphconv}, a cloned vertex is connected to exactly one original edge, and no two original edges share a cloned vertex as shown in Figure \ref{fig:epart}(b).
Therefore, we can get the total number of cloned vertices by doubling the total number of original edges.
If we evenly partition the vertices in the converted graph into $k$ clusters and make sure no original edge is cut, we can ensure the two endpoints of any original edge belong to the same partition, and thereby we can determine the corresponding original edge belongs to that partition, too.
In the meantime, we ensure that the number of original edges is balanced across all partitions, since every partition has the same number of vertices.
Formally,

\begin{definition}
Assume a clone-and-connect transformation t transforms graph $D = (V,E)$ to $D' =
(V',E')$, $D
\xmapsto{t} D'$, we perform vertex partition on $D'$ and obtain a
vertex partition solution $VP_{D'}$ such that no original edge is cut. From $VP_{D'}$, we can reconstruct an
edge partition solution of $D$, $EP_{D}$. We define a reconstruction process $m$
such that $VP_{D'} \xmapsto{m} EP_{D}$ as follows.
For every original edge $e'\in E'$ with $e' = (u',v')$ in $D'$, assume both $u'$ and
$v'$ fall into the i-th partition; then its corresponding edge $e \in D$, falls into the i-th partition as well.
\end{definition}

As shown in Figure \ref{fig:epart}(d), the edges $A$, $B$, $C$
are in the same cluster since their endpoints are all in the partition on the left.
Using this approach, we can map the vertex partition back to the edge partition
by checking every edge on which partition its endpoints fall into (shown in Figure \ref{fig:epart}(e)).

To ensure no original edge is cut in the vertex partition, we assign a very large weight to all original edges in the converted graph so that the vertex partition process will only cut auxiliary edges.
Also note that in the conversion process described in Definition \ref{def:graphconv},
there are several different ways to connect cloned vertices to form a path.
We choose to connect them in index order in practice.
Any other way to connect cloned vertices is fine, since auxiliary edges are used to represent data sharing relationship between original edges (tasks).

The balance degree of $D$'s edge partition is equal to that of $D'$'s vertex partition.
We find that existing vertex partition algorithms \cite{karypis1995metis} can produce rather balanced results.
The \emph{balance factor} is used to measure the balance degree of partition, which is calculated by dividing maximum partition size by average partition size \cite{karypis1995metis}.
It is typically less than 1.03 in practice.

We will show the mapping relationship between the optimum vertex partition and the optimum edge partition with rigorous analytical results in Section \ref{subsec:analytical}.

\subsection{Comparison with Other Methods}
\label{sect:tech:compare}

\begin{figure*}
	\centering
	\includegraphics[width=0.7\textwidth]{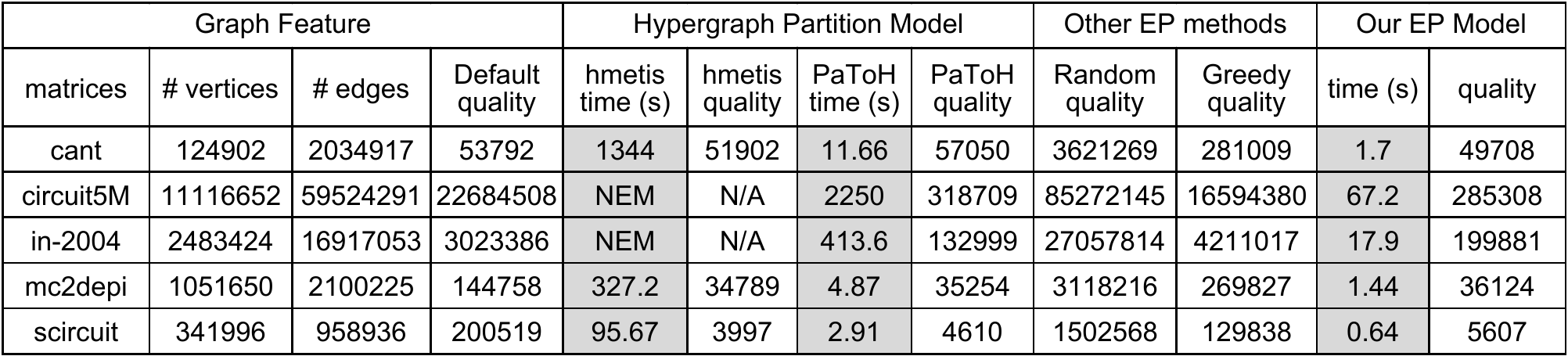}
	\caption{Our proposed EP model versus other partition methods. NEM represents
not-enough-memory errors.  }
	\label{fig:comparison}
\end{figure*}


Figure \ref{fig:comparison} compares our proposed conversion based edge partition model (EP model) with other existing task partition methods.
The experimental environment is introduced in Section \ref{sect:eval:env}.
For the vertex partition of the converted problem in our EP model, we use METIS \cite{karypis1995metis} library.
Five data-affinity graphs are constructed using representative input matrices
selected from the Florida matrix collection \cite{FloridaMat} and matrix market
\cite{MATRIXMARKET} (note that a sparse matrix usually represents a graph in or
is used in sparse matrix vector kernels).
We show the degree distribution for the data-affinity graphs in Figure
\ref{fig:4pdf}. We did not show the frequency of
degree for the \emph{mc2depi} data set, since its pattern is relatively simple,
with degree 2 of 5.70532e-04\% occurrence, degree 3 of 0.583654\% occurrence,
and degree 4 of 99.4158\% occurrence.  

The selected data-affinity have different degree
distribution patterns. The degree of the \emph{cant} data-affinity graph is
between 0 and 40. The \emph{circuit5M} has a more random degree distribution and
we only show part of the x range for readability.
The \emph{mc2depi} data set has nodes of degree 2 to 4, resembling the meshes
(at most 4 neighbour particles ) in computation fluid dynamics. Two
of the data sets exhibit the power-law pattern: the \emph{in$-$2004} data set and
the \emph{scircuit} distribution. We further show the log$-$log scale for these
two data sets in Figure \ref{fig:loglog2}. Regardless of the type of degree
distribution, our algorithm always outperforms the classical
\emph{vertex-centric} algorithm.

\begin{figure}
\centering
\includegraphics[width=0.4\textwidth]{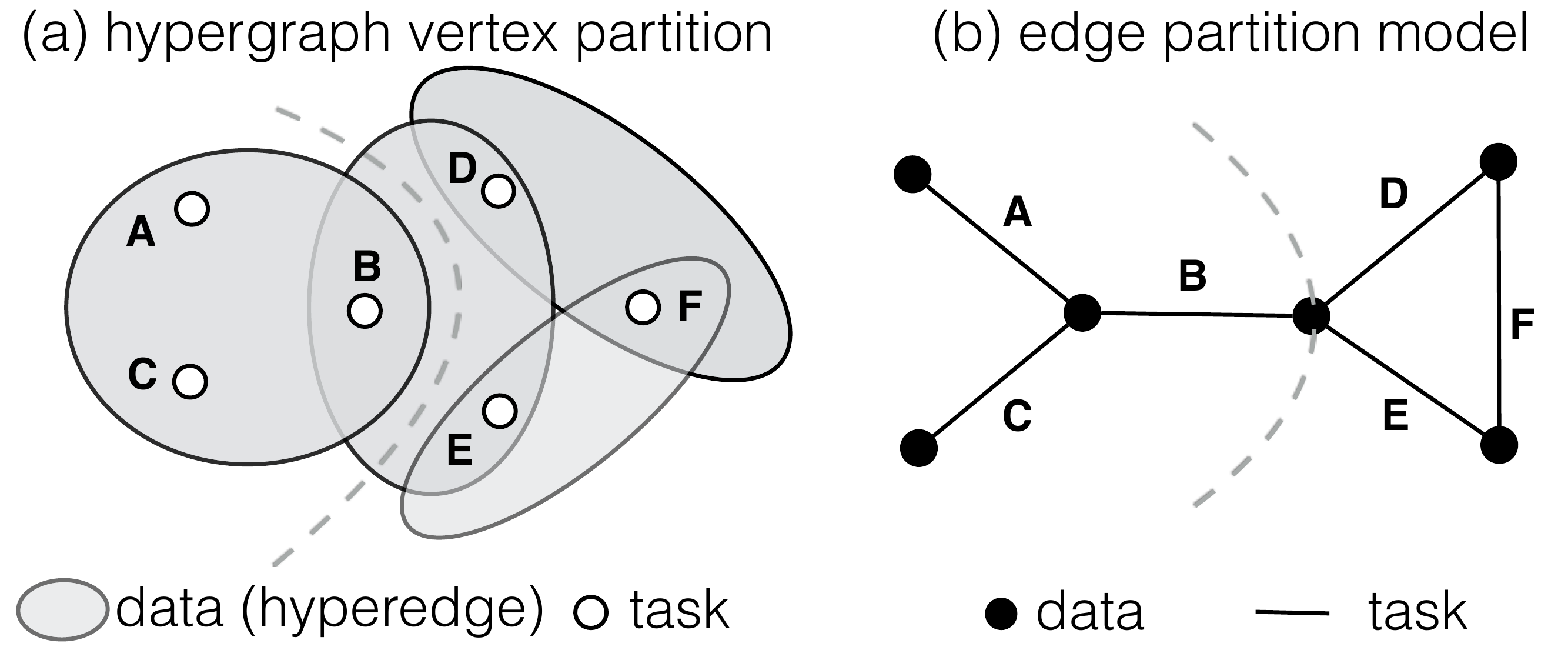}
\caption{Hypergraph model versus edge partition model.}
\label{fig:hyper}
\end{figure}

\paragraph{Comparison with hypergraph model \cite{Hendrickson+:PC00}}
A hypergraph is a special type of graph where an edge may connect more than two vertices.
Therefore an edge is also called a hyperedge.
In the hypergraph task partition model, unlike our EP model, a vertex represents a task, and a hyperedge represents a data object where it covers all vertices (tasks) that share that data object.
The goal of maximizing data sharing is equivalent to partitioning vertices (tasks) into $k$ clusters so that the number of times hyperedges are cut is minimized.
Figure \ref{fig:hyper} shows an example to compare the hypergraph model with the EP model.
We also show the optimum partition for both models.
In Figure \ref{fig:hyper}(a) one hyperedge is cut, which corresponds to the one cut vertex in Figure \ref{fig:hyper}(b).

For the hypergraph partition, we use hMETIS \cite{karypis1998hmetis}, a multilevel hypergraph partition tool, and PaToH \cite{catalyurek1999patoh}, the fastest hypergraph partition implementation we are aware of.

From Figure \ref{fig:comparison}, we see that PaToH is faster than hMETIS in the hypergraph model, and our basic EP model is significantly faster than both of them in all cases.
The partition quality, measured as the data reuse cost in Definition \ref{def:epproblem}, shows that our EP model generates similar quality of partitions as PaToH and hMETIS.
The quality of both the hypergraph model and our EP model are better that of the default task scheduling shown in the ``default quality'' column, which indicates that data sharing within each thread block is improved significantly after partitioning.
Overall, our basic EP model is faster than the fastest hypergraph partition model and yet produces high quality results.

Another important thing to notice is that our EP model scales better compared to the hypergraph model with respect to graph size.
For small graphs such as \emph{scircuit}, our EP model is about 4 times faster,
while for large graphs such as \emph{circuit5M} and \emph{in-2004}, the EP model is tens of times faster.

\paragraph{Comparison with other edge partition methods}
PowerGraph \cite{gonzalez2012powergraph} proposes two edge partition based task scheduling methods.
Both methods go through all edges linearly to distribute them among partitions.
The random based method randomly assigns edges into partitions.
The greedy based method prioritizes choosing partitions that already possess the endpoints of the to-be-assigned edge.
If no such partition is found, then the partition with the fewest edges is selected to ensure balance.
Figure \ref{fig:comparison} shows the partition quality of both methods.
We do not show the partition time since both methods are expected to be fast.
Compared with hypergraph and our EP models, both of them have significantly worse partition quality because of their random or greedy nature.
In many cases, their quality is worse than that of the default scheduling.
In such scenarios, rescheduling tasks only degrades the performance.
Therefore, we conclude that these two partition methods are not applicable for complicated data sharing patterns exhibited in GPU programs.

\subsection{Approximation Bounds}
\label{subsec:analytical}

Our EP model is not only good in practice, but also strong in theory. We provide the proof for approximation factor between
$\mathcal{O}(\sqrt{\log{}m\log{}k})$ and
$(d_{max}-1)\mathcal{O}(\sqrt{\log{}m\log{}k})$. We describe it in two steps.

\begin{theorem}
With the clone-and-connect transformation t such that $D \xmapsto{t} D'$
where $D=(V,E)$ and $D'=(V',E')$. The edge cut cost of the vertex partition of
$D'$ is greater than or equal to the vertex cut cost (in Equation \ref{commcost}) of
the corresponding edge partition of $D$, $C_{vp}(D') \geq C_{ep}(D)$.
\label{lm:costconv}
\end{theorem}

First recall that in Section \ref{subsec:alg}, we have ensured in the
vertex partition of $D'$ that only auxiliary edges are cut. For every set of cloned
vertices $v'_i$, i $= 1...d$ in $D'$ that corresponds to the vertex $v$ of degree $d$
in $D$, they are connected with $d-1$ auxiliary edges to form a path. If $l$ out
of the $d-1$ auxiliary edges are cut, the total number of distinct edge
clusters $v$ fall into in $D$, is at most $l+1$. Therefore, the
cut cost of $v$ in the edge partition of $D$ is at most $l$. When
converting the vertex partition of $D'$ into the edge partition of $D$, we sum
up the total vertex cut cost and thus the
total number of auxiliary edges cut in $D'$ is greater than or equal to the vertex
cut cost in $D$.

\begin{theorem}
There exists a clone-and-connect transformation t, $D \xmapsto{t} D'$ such
that the edge cut cost of the optimal vertex partition of $D'$ is the same as
the vertex cut cost of the optimal edge partition of $D$. For all legal
clone-and-connect transformations $t \in T$, in the worst case, the
edge cut cost of the optimal vertex partition of $D'$ is $d_{max}-1$ times of
the vertex cut cost of the optimal edge partition of $D$, while $d_{max}$
represents the maximum degree of the graph $D$.
\label{thrm:bound}
\end{theorem}


The first task is to prove that there exists a clone-and-connect
transformation $t$ such that the transformed vertex partition problem is the
{\em dual} of the original edge partition problem. We construct such a
transformation using an optimal edge partition $EP_{opt}$ on the graph $D=(V,E)$. Using the
optimal edge partition solution, we know which edges are grouped to the same
partition. Then we determine the transformation $t$ as follows. First, the cloning
process is the same as in Definition \ref{def:graphconv}. In the connecting
phase, for every set of cloned vertices $v'_i$, $i = 1...d$ that connect to every
original edge, we first divide each set into $k$ groups with respect to which
partition their incident original edges belong to. Within every set of cloned vertices,
within every group, we connect the
cloned vertices to form a path, then we connect all $k$ groups into
one path. Such a transformation $t$ is constructed. It is easy to see that if we reverse this path
connecting process, we can obtain a balanced vertex partition of $D'$ that can
be mapped to the optimum edge partition solution of $D$, which is $EP_{opt}$ we
defined earlier.

Since we do not \emph{a priori} know the optimal edge partition of the
original graph $D$, when connecting the cloned vertices into a path, we connect
them arbitrarily. Assuming the converted graph with a priori knowledge of
optimal partition of $D$
is $D'_{opt}$, the second task is to prove that with an arbitrary connecting
process, the edge cut cost of the optimum vertex partition of the converted
graph $D'$ is at most
$d_{max}-1$ times of the edge cut cost of the vertex partition of $D'_{opt}$
since $D'_{opt}$ maps directly back to $D_{opt}$ ($D_{opt}$ is the optimal edge
partition of the original problem).
This step is easy to prove as well, since for every set of $d$ cloned vertices that
correspond to the cut vertex in the optimal edge partition solution of $D$, at
most all auxiliary edges are cut ($d-1$ cuts) and in the best case, at least one
auxiliary edge is cut, the edge cut cost in the vertex partition of $D'$ is thus
at most $d_{max}-1$ the edge cut cost in the vertex partition of $D'_{opt}$.
According to Theorem \ref{lm:costconv}, in the reconstructed edge partition of
$D$ from
the vertex partition of $D'$, the vertex cut cost of $D$ is less than or equal
to the
edge cut cost of $D'$, $C_{ep}(D) \leq C_{vp}(D')$. Thus the reconstructed edge
partition solution from the optimum vertex partition of $D'$ is at most
$d_{max}-1$ times the optimum solution of edge partition of $D$.



The vertex partition problem can be approximated with a factor of $\mathcal{O}(\sqrt{\log{}n\log{}k})$ for a graph with $n$
vertices and $k$ partitions \cite{Krauthgamer+:SODA09}. Therefore we can approximate our solution by a
factor of $(d_{max}-1)\mathcal{O}(\sqrt{\log{}m\log{}k})$ since our converted
graph has $2m$ vertices where $m$ is the number of edges in the original graph.

%% file: practical_new.tex
\begin{figure}[htb]
	\centering
	\includegraphics[width=0.8\linewidth]{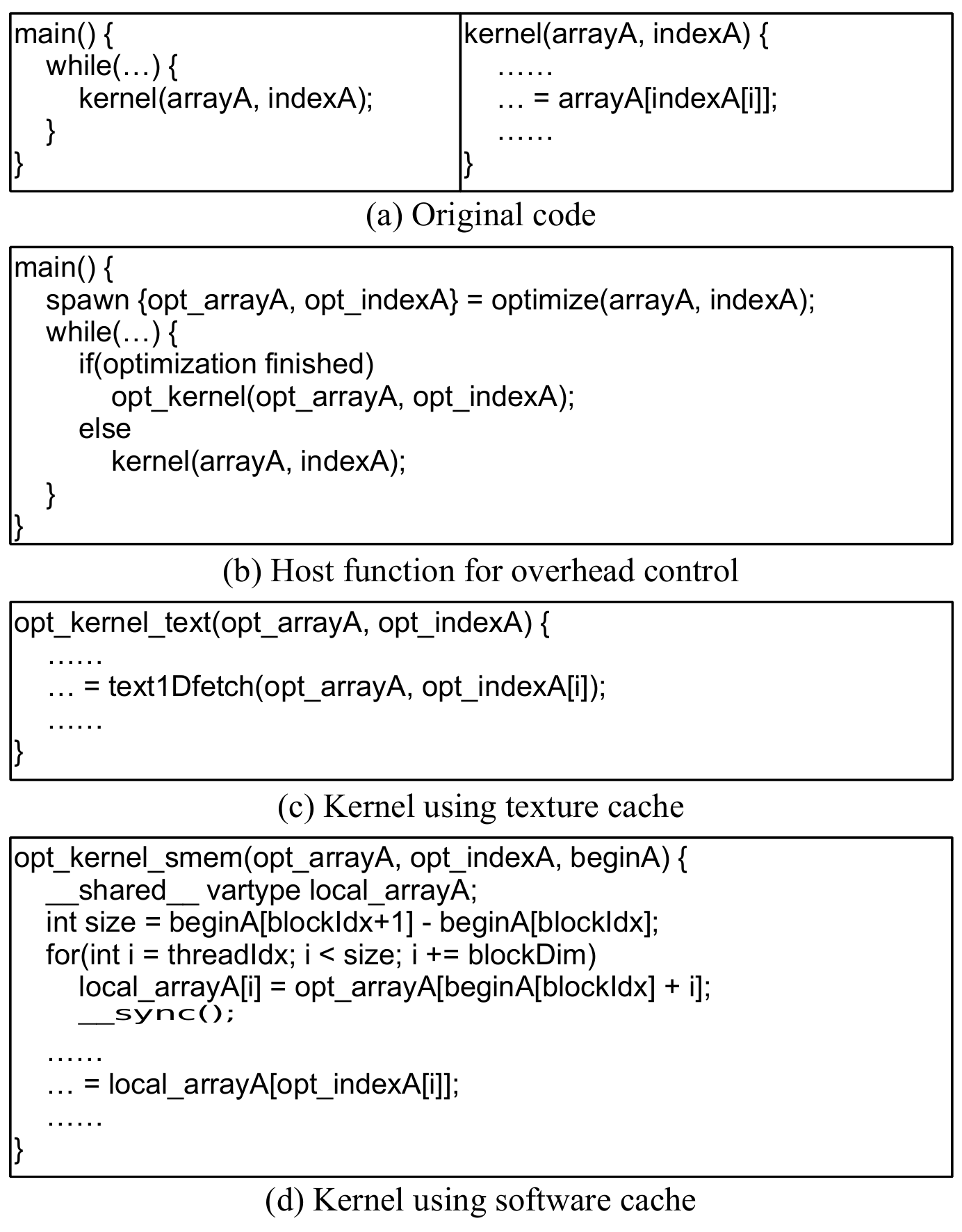}
	\caption{Example of code transformation.}
	\label{fig:code_trans}
\end{figure}

\section{Program Transformation}
\label{sec:appl}

In this section, we will introduce how to apply the partition result of our EP model to GPU programs for data sharing optimization.
Figure \ref{fig:code_trans} shows an example of original and transformed code that enables data sharing optimization.

For the host function on the CPU side in Figure \ref{fig:code_trans}(b), it first spawns a separate CPU thread to perform data sharing optimization.
The optimization process will be introduced in Section \ref{sect:appl:process}.
We use a separate thread for optimization to prevent it from adversely affecting the performance of the main program.
Assume the GPU kernel is called multiple times.
Before every kernel call, we check if the optimization process has been completed.
If so, the optimized kernel will be called with the optimized input.
Otherwise, the original kernel is called as usual.
The details of how we control optimization overhead, as well as the situation when the GPU kernel is only called once, will be discussed in Section \ref{sect:appl:overhead}.

If the target cache for data sharing optimization is texture cache,
the host function binds the optimized data layout to texture memory using the CUDA built-in function {\em cudaBindTexture()}.
The GPU kernel prefixes every data reference using {\em tex1Dfetch()} as shown in Figure \ref{fig:code_trans}(c) so that referenced data will be cached automatically by texture cache.

If the target cache is software cache, the kernel function on the GPU side in Figure \ref{fig:code_trans}(d) requires three major modifications in order to leverage optimized input.
First, an array local\_arrayA is created in software cache to store shared data.
Then, each thread block loads its shared parts from the input array (opt\_arrayA) into the local\_arrayA.
To minimize memory divergence \cite{Zhang+:ASPLOS11, Wu+:PPoPP13}, memory
accesses are coalesced into as few contiguous memory segments as possible. 
At last, we replace the reference of the original input array with that of the local array which resides in software cache.

\subsection{Optimization Process}
\label{sect:appl:process}

\begin{figure}[htb]
	\centering
	\includegraphics[width=\linewidth]{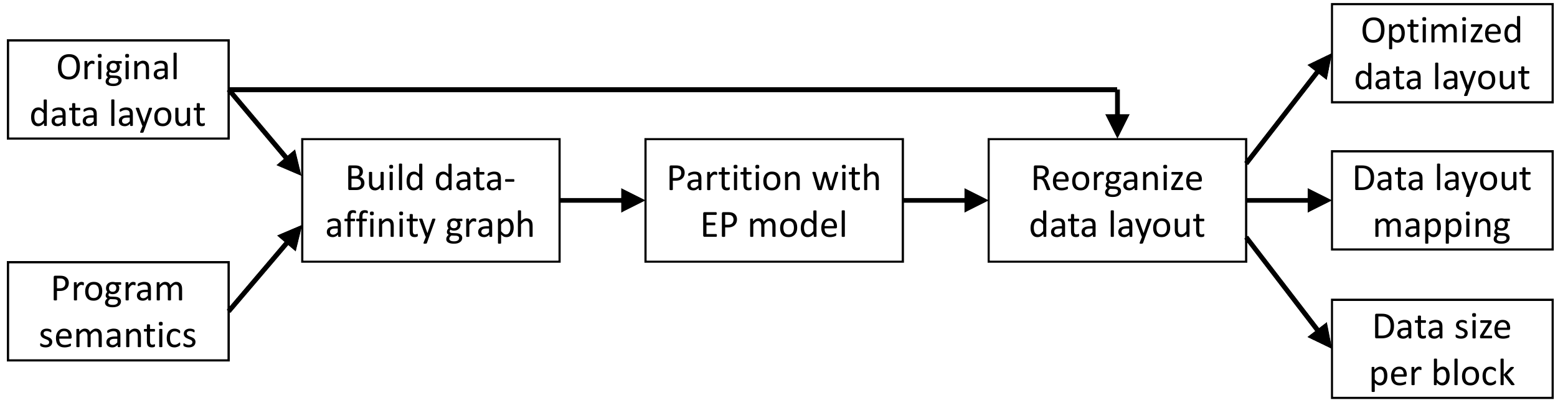}
	\caption{Optimization workflow.}
	\label{fig:workprocess}
\end{figure}

Figure \ref{fig:workprocess} illustrates the working process of data sharing optimization.
The data access pattern of a GPU program is determined by both program semantics and input data.
Therefore, we extract the data access pattern to build the data-affinity graph at first.

After data-affinity graph is built, we examine the graph before actually partitioning it using EP model.
First, we check if there is enough data reuse by checking the degree frequency
of the data-affinity graph.
If there is little data reuse,  then we omit the partition process and use the original input.
Secondly, we check if the graph follows a special pattern such as clique, path, complete bipartite, etc.
For these special graphs, we have a preset optimal partitioning schedule using
the EP model offline, and thus we use the preset partition pattern. 

If we determine that task partitioning/reschedule is necessary for the generated
data-affinity graph, we apply the EP algorithm to perform task partitioning as described in Section \ref{sec:tech}.

At last, we reorganize tasks among thread blocks based on the partition result
and we reorganize data layout as well. We use the cpack algorithm
\cite{Ding+:PLDI99}
to perform data layout transformation based on our task schedule. 
After the data layout is transformed (from arrayA to opt\_arrayA in Figure
\ref{fig:code_trans}), the data references need to be updated as well and the
new index array (opt\_indexA in Figure \ref{fig:code_trans}) is passed to the optimized kernel.


\subsection{Adaptive Overhead Control}
\label{sect:appl:overhead}

To reduce overhead at runtime, we perform data sharing optimization using a separate thread on the CPU while kernel is executed on the GPU, so that the optimization process can run asynchronously.
We utilize a CPU-GPU pipelining technique to overlap computation and optimization for GPU programs \cite{Zhang+:ASPLOS11, Wu+:PPoPP13}.

In real GPU programs, a kernel is usually called within a loop, especially in scientific computation programs.
In such scenarios, We check if the asynchronous optimization is completed before calling the kernel and apply the optimization if so.
If the optimization thread does not complete when the program finishes, we terminate it to guarantee no slowdown.


If a kernel is only called once, we perform \emph{kernel splitting} \cite{Zhang+:ASPLOS11} which breaks one execution of a kernel into multiple executions of the same kernel with a smaller number of threads.
Then we can apply the asynchronous optimization method.

It is, however, possible that the transformed kernel runs slower than the original kernel. To prevent such cases from degrading performance, we record the transformed kernel runtime the first time it runs, and compare it with the original kernel runtime.
If the first run of the transformed kernel is slower, then we fall back to the original kernel in the next iteration.

%% file: eval_new.tex
\section{Evaluation}
\label{sec:eval}

\subsection{Experiment Setup}
\label{sect:eval:env}

Our hardware platform is a machine with an NVIDIA GeForce GTX680 GPU with CUDA computing capability 3.0. It includes 8 streaming multiprocessors (SMs) with
192 cores on each of them and 1536 cores in total. On each SM there is 64 KB of L1 cache and shared
memory (i.e., software cache), shared by its 192 cores. There are three size configurations of L1
cache and shared memory: 16KB/48KB, 32KB/32KB and 48KB/16KB. Since L1 cache in
GTX680 is allowed for local memory data only, we always configure shared memory to be 48KB.
There is also 48KB texture cache per SM.
The CPU is Intel Core i7-4790 at 3.6 GHz with 8MB last-level cache.

\begin{table*}[htb]
\footnotesize
\centering
\begin{tabular}{|l|c|c|c|c|}
\hline
Benchmark & Application Domain & Input & Cache type \\
\hline
b+tree \cite{Che+:IISWC09} & Tree data structure &  One-million size database
query & Software\\
bfs \cite{Che+:IISWC09}  & Breadth first search  & One-million node graph & Texture\\
cfd \cite{Che+:IISWC09}  & Computation fluid dynamics & 97K, 193K, and 0.2M node
meshes & Software\\
gaussian \cite{Che+:IISWC09}  & Gaussian elimination & Linear system with 1024
unknown variables & Software  \\
particlefilter \cite{Che+:IISWC09}  & SMC for posterio density estimation &
Tracking of 1000 particles & Software \\
streamcluster \cite{Che+:IISWC09}  & Data stream clustering & 65,536 data
points & Software \\
CG \cite{CUSP} & Sparse matrix multiplication & Various size input matrices & Both\\
\hline
\end{tabular}
\caption{Benchmark summary.}
\label{tbl:bench}
\end{table*}

We use seven applications from various domains including data
mining, computation biology, object tracking, scientific simulation, and graph
traversal. This set of benchmarks are representative of important contemporary
workloads that can benefit from cache locality enhancement. The benchmarks are listed in Table \ref{tbl:bench}.

Six applications are from the Rodinia benchmark suite \cite{Che+:IISWC09}.
We also evaluate an important kernel, the sparse matrix vector multiplication
(SPMV) kernel, which is widely used in different types of applications including
machine learning and numerical analysis. The SPMV kernel is also frequently used for evaluating graph partition models on parallel CPUs \cite{karypis1999parallel, karypis2000multilevel}.
We report the performance of SPMV kernel and the task scheduling overhead. We also evaluate the asynchronous
optimization method.  We run the SPMV kernel in
the context of the conjugate gradient (\emph{CG}) \cite{CUSP} application. The
CG application calls SPMV iteratively until the solution converges. 
All benchmarks are compiled on a system that runs 64-bit Linux with kernel version 3.1.10 and CUDA 5.5.

We use the input sets bundled with the benchmark suite or real-world input sets. For example, the \emph{cfd} benchmark from Rodinia benchmark suite \cite{Che+:IISWC09} has three input meshes: \emph{fvcorr.domn.097K}, \emph{fvcorr.domn.193K} and \emph{missile.domn.0.2M}.
The sparse matrix inputs to \emph{CG} come from the University of Florida sparse matrix collection \cite{FloridaMat} and matrix market \cite{MATRIXMARKET}.

In the rest of this section, we first evaluate the performance of sparse matrix
vector kernel over a large number of sparse matrices and various configurations.
Then we show the evaluation results of the six applications from Rodinia.

\subsection{Sparse Matrix Workloads}

Conjugate gradient (CG) method \cite{Hestenes+:NBS52} is an important algorithm for solving linear
systems, whose matrices are usually large and sparse. The conjugate gradient
method is an iterative process that repeated invokes sparse matrix vector
multiplication until the result converges. We focus on the total amount of time
for sparse matrix vector multiplication in CG in the following
discussion.

We partition the sparse matrix workload with our edge partition model and hypergraph model to
maximize data reuse and compare their performance with the popular SPMV implementation from CUSP \cite{CUSP} and CUSPARSE \cite{CUSPARSE}.
The CUSP spmv kernel orders the data layout such that all non-zero elements are sorted by row indices and then it distributes the non-zero elements evenly to threads.
We are not aware of CUSPARSE's task distribution scheme since it is not open source, however since it is faster than CUSP for most inputs, we include CUSPARSE for comparison.

In the data-affinity graph of SPMV, there is a vertex for every element in the input vector $x$ and output vector $y$.
For each non-zero element $A[i,j]$ in the input matrix $A$, an edge is added to connect vertex $j$ in the input vector and vertex $i$ in the output vector, since a non-zero $A[i,j]$ implies a multiplication with $x_j$ to get $y_i$.
The data-affinity graph of SPMV is naturally a bipartite graph.
With the data-affinity graph, we perform edge partition and let one thread block be mapped to one partition.
We use software cache for the corresponding shared input and output vector elements within each thread block.
We will also show the results of using texture cache later.

\begin{table*}[t]
\centering
\footnotesize
\begin{tabular}{|l|c|c|c|c|c|c|c|} \hline
Name & Dimension & Nnz & CUSPARSE time & EP time & EP partition & HP time & HP partition \\
\hline
cant & 62K*62K & 2.0M & 2.53 & 2.89 & 1.702 & 2.92 & 11.66 \\
circuit5M & 5.6M*5.6M & 59.5M & 21599 & 783.2 & 67.157 & 784.2 & 2250 \\
cop20k\_A & 121K*121K & 1.4M & 25.93 & 20.17 & 1.457 & 19.99 & 5.76 \\
Ga41As41H72 & 268K*268K & 9.4M & 19.37 & 15.29 & 17.922 & 15.26 & 194.5 \\
in-2004 & 1.4M*1.4M & 16.9M & 430.9 & 359.4 & 17.889 & 355.7 & 413.6 \\
mac\_econ\_fwd500 & 207K*207K & 1.3M & 31.54 & 18.29 & 1.342 & 18.14 & 5.04 \\
mc2depi & 526K*526K & 2.1M & 36.45 & 28.31 & 1.436 & 28.36 & 4.87 \\
scircuit & 171K*171K & 0.96M & 20.42 & 13.62 & 0.642 & 13.51 & 2.91 \\
\hline
\end{tabular}
\caption{Matrix Information.}
\label{tbl:matrices}
\end{table*}

\begin{figure}
\centering
\includegraphics[width=0.8\linewidth]{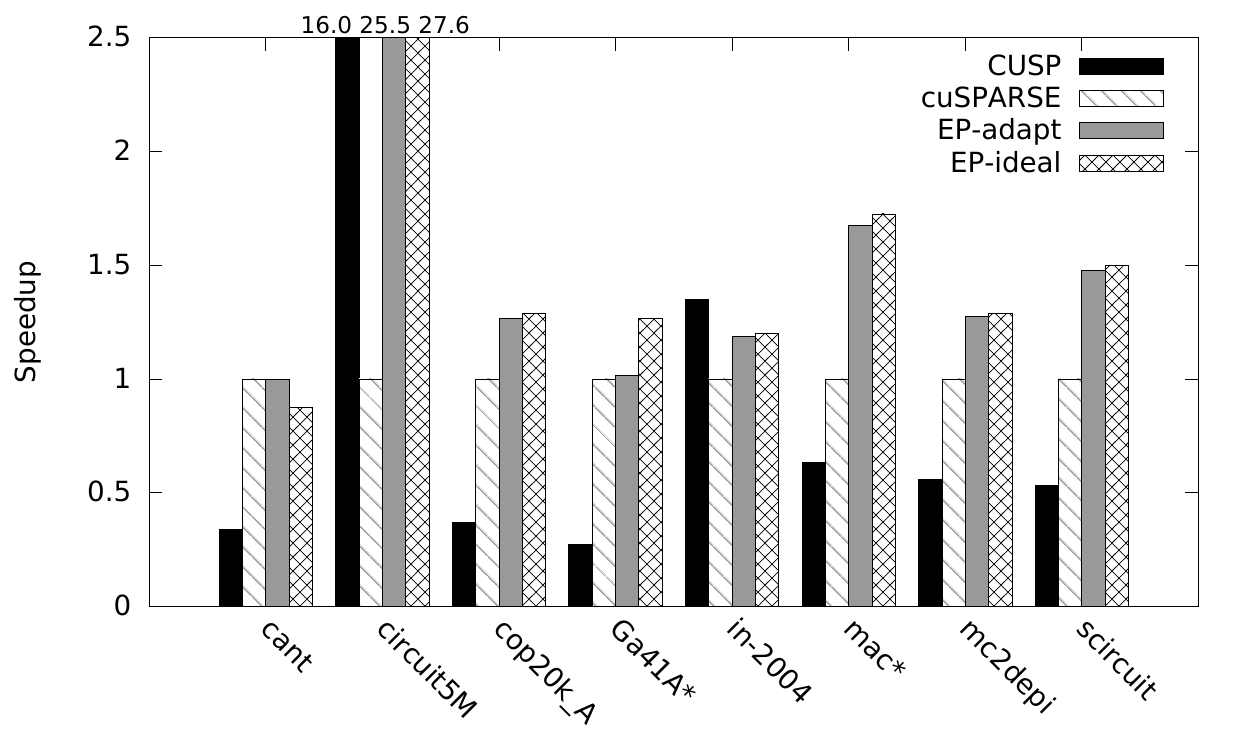}
\caption{Speedup of various spmv kernels.}
\label{fig:spmv_speedup}
\end{figure}

We use 8 matrices with different data access patterns to show the applicability of our EP model.
These matrices are shown in Table \ref{tbl:matrices}.
\emph{Nnz} represents the total number of nonzero elements.
\emph{CUSPARSE time} column shows total SPMV kernel execution time of CUSPARSE.
\emph{EP time} and \emph{HP time} show the total kernel execution time of our EP model and hypergraph model respectively.
\emph{EP partition} and \emph{HP partition} show their partition time.
All times are measured in seconds.

In Table \ref{tbl:matrices}, we first observe similar kernel execution time for
EP and HP models, which confirms our EP model can produce similar partition
quality compared to hypergraph partition model. (Note that in Table
\ref{tbl:matrices} we show the partition time and kernel execution time without
applying the asynchronous optimization method -- the overhead control. We will
show the results with overhead control later.)
On the other hand, hypergraph partitioning incurs much larger partition overhead compared to our EP model.
Its partition time occupies 205\% of the total kernel time of CUSPARSE on average, which means the hypergraph partition model could not finish optimization for most matrices before program ends.
On the other hand, the partition time of our EP model only occupies 22.7\% of total time on average.
Therefore, hypergraph partitioning is infeasible for GPU applications due to its large overhead.
Other edge partition methods have been proved to have worse quality than the default scheduling in Figure \ref{fig:comparison}.
Thus we focus on the study of our EP model in the rest of this paper.

Figure \ref{fig:spmv_speedup} compares the performance of four versions of SPMV kernel execution, including CUSPARSE, CUSP, our EP model that does not consider optimization overhead (EP-ideal), and the one that takes overhead into consideration (EP-adapt).
We set the thread block size to be 1024. We use
CUSPARSE kernel time as the baseline, since it is faster than CUSP for most matrices.

In Figure \ref{fig:spmv_speedup}, we can see that CUSP is slower than
CUSPARSE except in two cases, \emph{circuit5M} and \emph{in-2004}.
However, our EP model based approach is faster than both of them except for \emph{cant}, where the quality of default scheduling is similar to that of our EP model as shown in Figure \ref{fig:comparison}.
When using EP-adapt for \emph{cant}, there is almost no slowdown for \emph{cant}. That is because adaptive overhead control checks if optimization is beneficial and if not it would stop using the optimized kernel.
EP model is slightly worse than CUSP for \emph{in-2004} because there is very limited data reuse, which causes EP model to use a large amount of software cache, adversely affecting occupancy.
We also observe that in most cases, the performance of EP-adapt is similar to that of EP-ideal except for \emph{Ga41As41H72}, where the partition time occupies most kernel execution time and we can only optimize a small portion of SPMV invocations.

\begin{figure}
\centering
\includegraphics[width=0.8\linewidth]{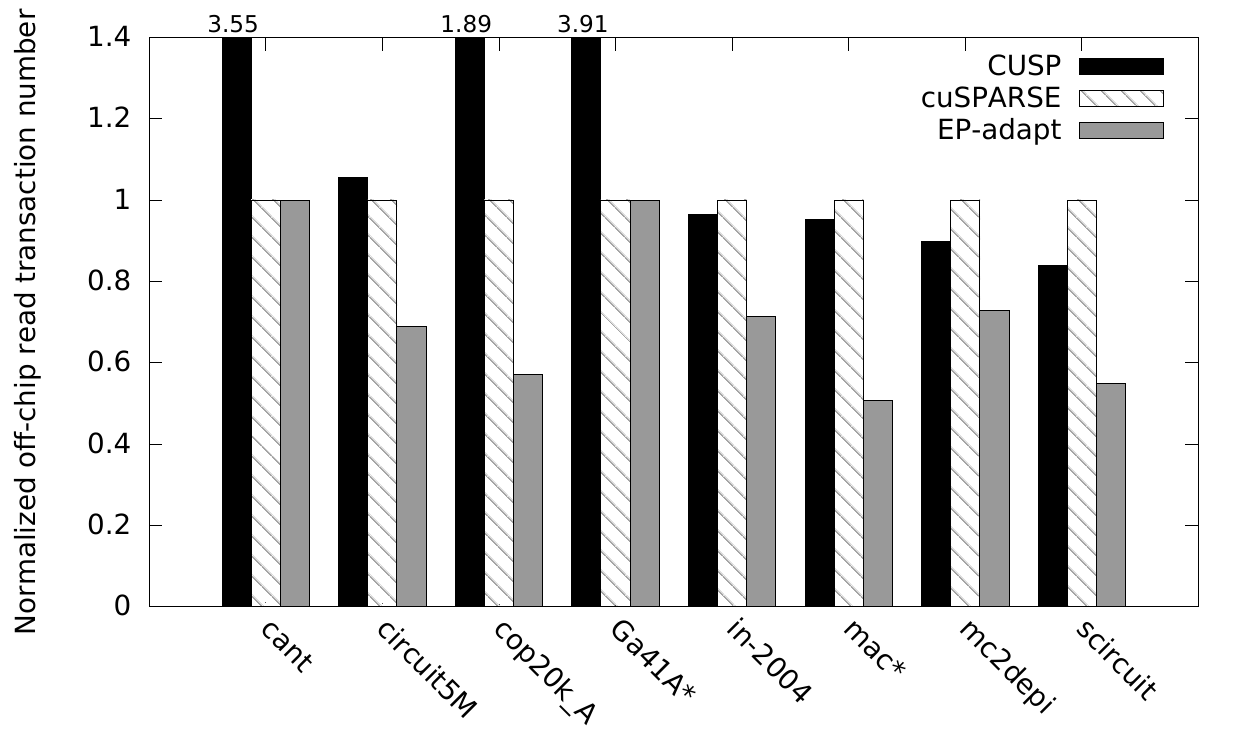}
\caption{Normalized transaction number of spmv.}
\label{fig:spmv_trans}
\end{figure}

Figure \ref{fig:spmv_trans} shows the normalized memory transaction number for three spmv kernels.
All results are normalized to that of CUSPARSE.
We observe that memory transaction number is reduced significantly for all matrices except \emph{cant} and \emph{Ga41As41H72}.
There is no memory traffic reduction for \emph{cant} and \emph{Ga41As41H72} due to the same reasons we have explained above.
Overall, the transaction reduction maps well to the performance results.

\begin{figure}
\centering
\includegraphics[width=0.8\linewidth]{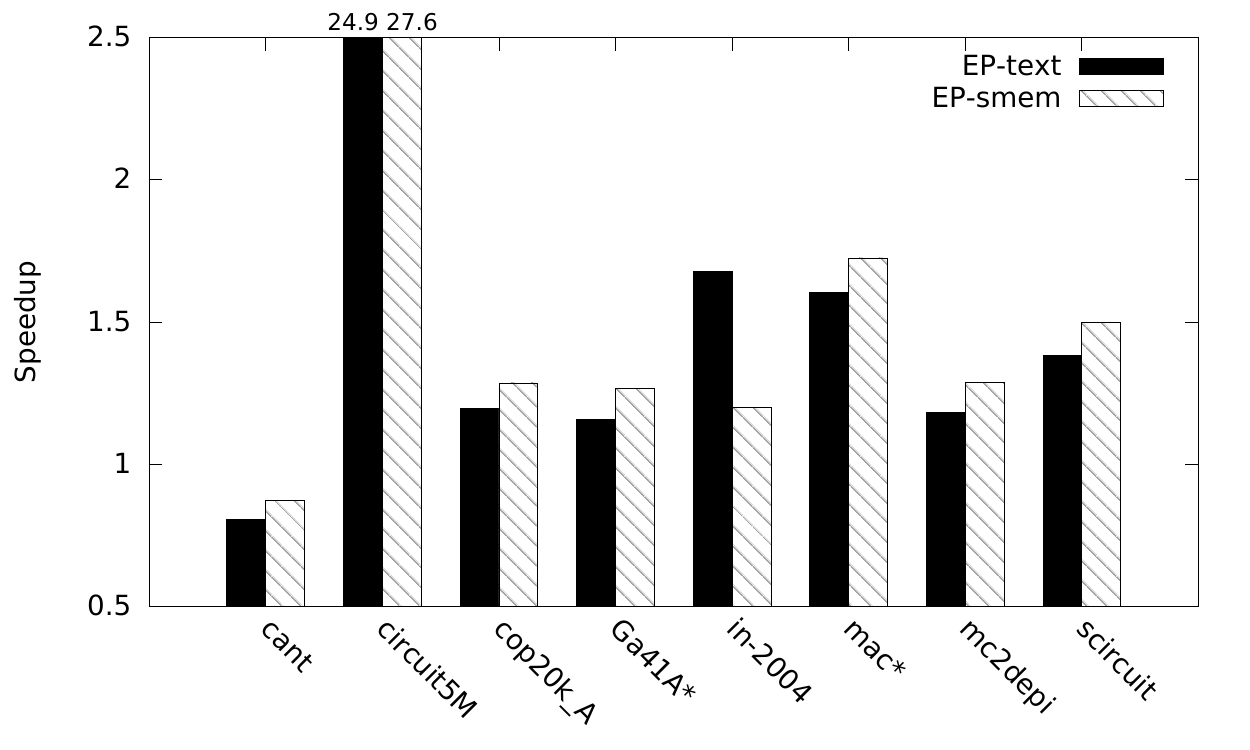}
\caption{Texture cache v.s. software cache.}
\label{fig:spmv_speedup_tc}
\end{figure}

\paragraph{Software cache v.s. texture cache}
Here we study the performance difference when using software and texture caches to optimize the data sharing of input vector in SPMV.
Since the output vector is write-shared, texture cache cannot be used to store it.
Figure \ref{fig:spmv_speedup_tc} compares their performance; EP-text represents the texture cache version and EP-smem represents the software cache (i.e., shared memory) version.
Again, the baseline is the performance of CUSPARSE.
Software cache version outperforms texture cache version for almost all matrices except \emph{in-2004}, in which case the large shared memory usage of EP-smem degrades thread level parallelism significantly as stated above.
Note that EP-text outperforms CUSP in Figure \ref{fig:spmv_speedup} where \emph{in-2004} is the only case our optimized kernel using software cache is not as good as CUSP.
Compared to CUSP and CUSPARSE, EP-text also has better performance.
In general, the texture cache based approach could potentially pollute cache by evicting data before it gets fully reused, while using software-managed cache does not have such a problem.
On the other hand, using texture cache can reduce programming overhead since it requires less effort for program and data transformation, and there is still sufficient performance improvement.
Thus there is a trade-off between programmability and performance.
In practice, we can choose either version based on programmability and performance preference.

\begin{table}
\centering
\footnotesize
\begin{tabular}{|l|c|c|c|c|c|c|} \hline
Block size	& \multicolumn{2}{c|}{256}	& \multicolumn{2}{c|}{512}	&
\multicolumn{2}{c|}{1024} \\ \hline
time (s) &		tex	& smem	& tex	& smem &	tex &	smem
\\\hline\hline
cant		&	310 &	279 &	313 &	289 &	330 &	305 \\
circuit5M		&	8312	& 7349	&  8524	& 7531	& 8691	& 7831\\
cop20k\_A		&	208	& 186	& 211	& 190	& 216	& 201 \\
Ga41A*		&	1353 &	1196 &	1378	& 1225	& 1409	& 1288 \\
in-2004		&	2471	&  4472	 &   2492	& 3490	& 2565	&  3594 \\
mac*		&	190	&  174	& 191.4	& 175	& 196	& 182 \\
mc2depi		&	332	& 284	& 307	& 277 &	307	& 283 \\
scircuit		&	143	& 130 &	144	& 133	& 147	& 136 \\ \hline
\end{tabular}
\caption{Performance on different thread block sizes.}
\label{tbl:block}
\end{table}

\paragraph{Thread block size} Finally, we show the sensitivity of our graph partition approach with respect to different thread block sizes.
Table \ref{tbl:block} shows the performance of our EP model
without overhead control (EP-ideal) under different thread block sizes for both software and texture caches.
At every thread block size, the performance difference between software and texture caches is similar to the observation we made earlier.
The results suggest for all shared memory kernels, the performance at small thread block size is slightly better than (or the same as) that of the larger thread block size except \emph{in-2004}, where shared memory version generally performs badly because of limited data sharing and low concurrency.
However, smaller block size implies larger block number, and thus longer partition time of EP model.
Taking both kernel performance and partition overhead into consideration, the performance using smaller block size is similar to that using large block size.
Therefore, we still choose 1024 as the default block size in our experiments.

\begin{figure*} [htb]
	\centering
	\subfloat[bfs]{\includegraphics[width=0.25\textwidth]{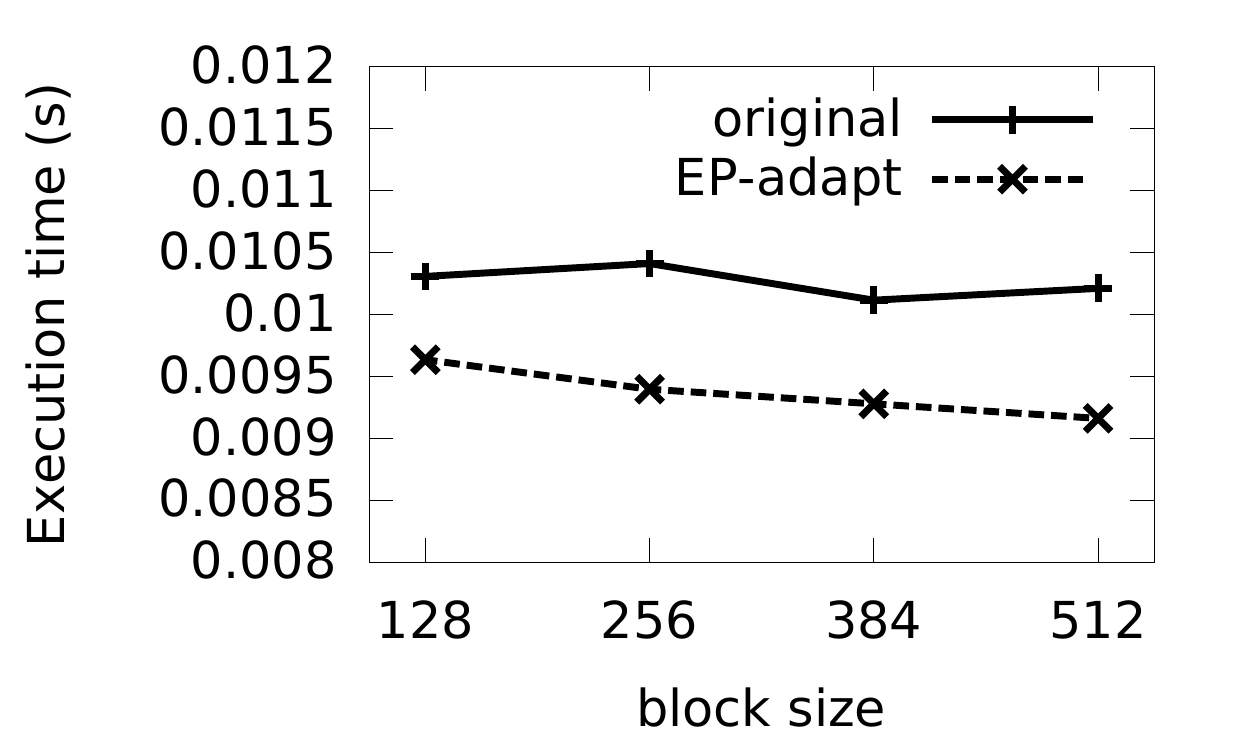}}
	\subfloat[b+tree]{\includegraphics[width=0.25\textwidth]{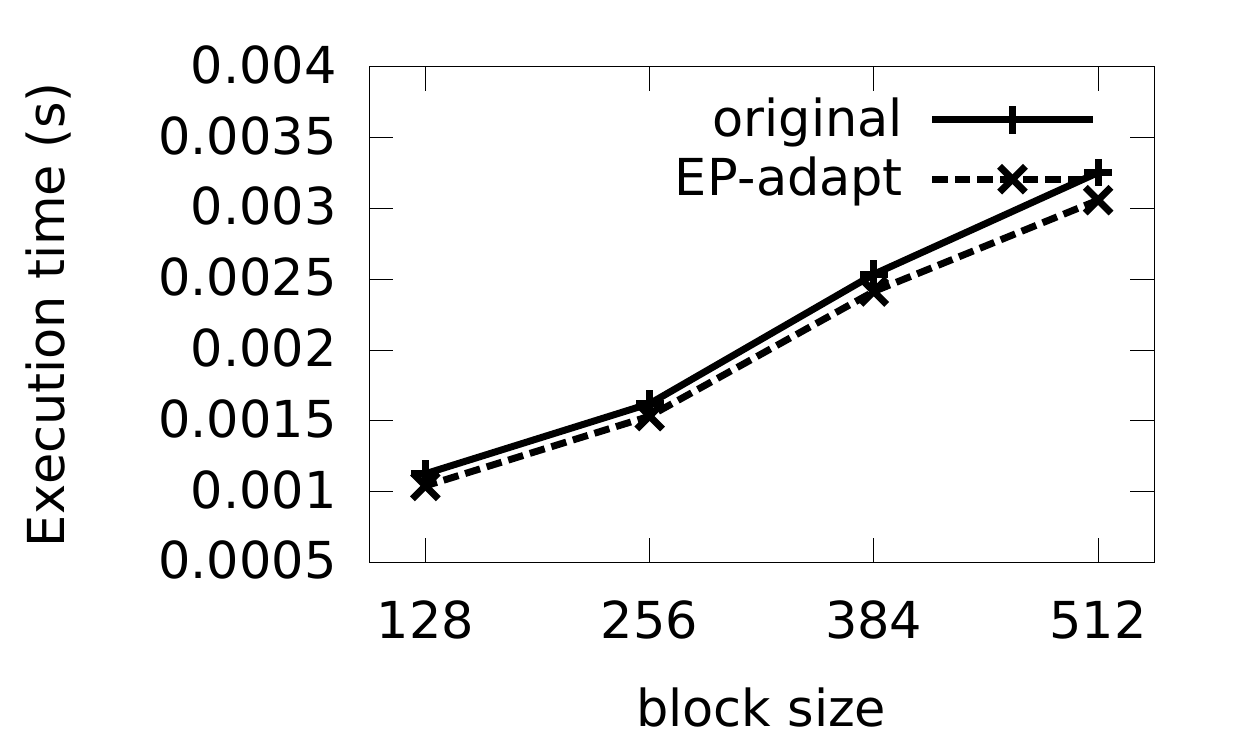}}
	\subfloat[cfd (f097K)]{\includegraphics[width=0.25\textwidth]{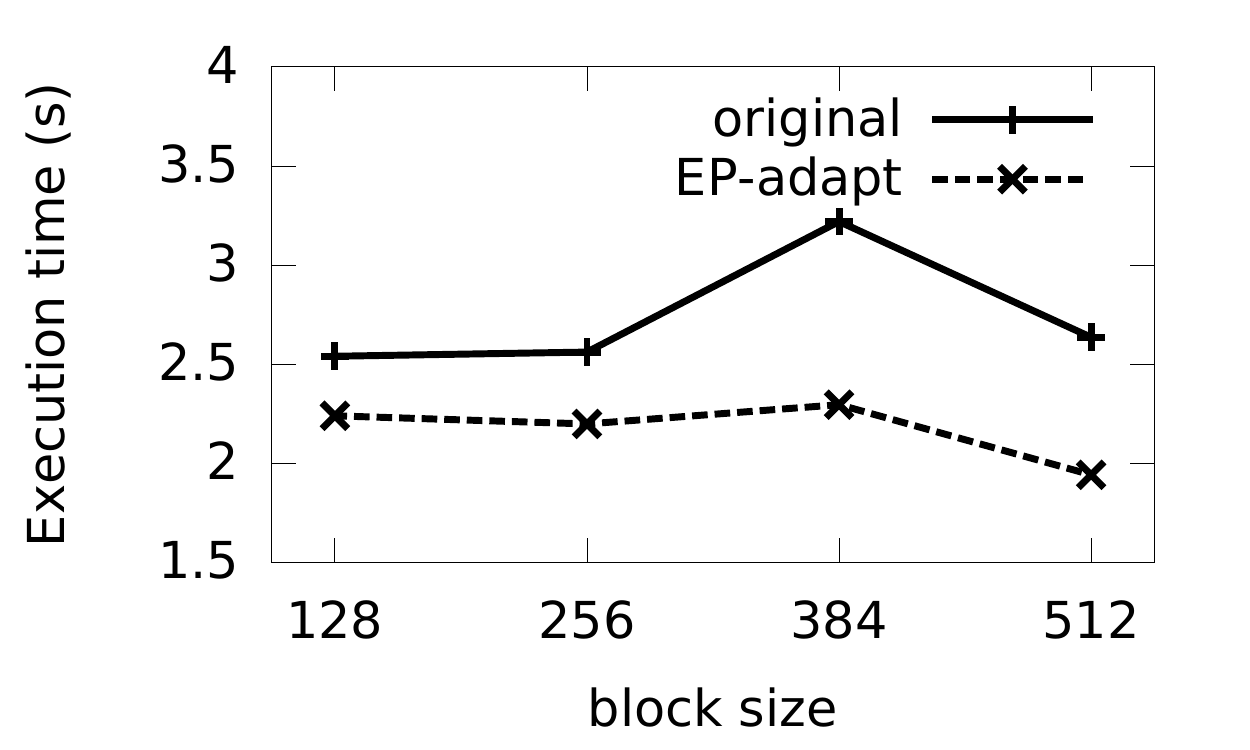}}
	\subfloat[cfd (f193K)]{\includegraphics[width=0.25\textwidth]{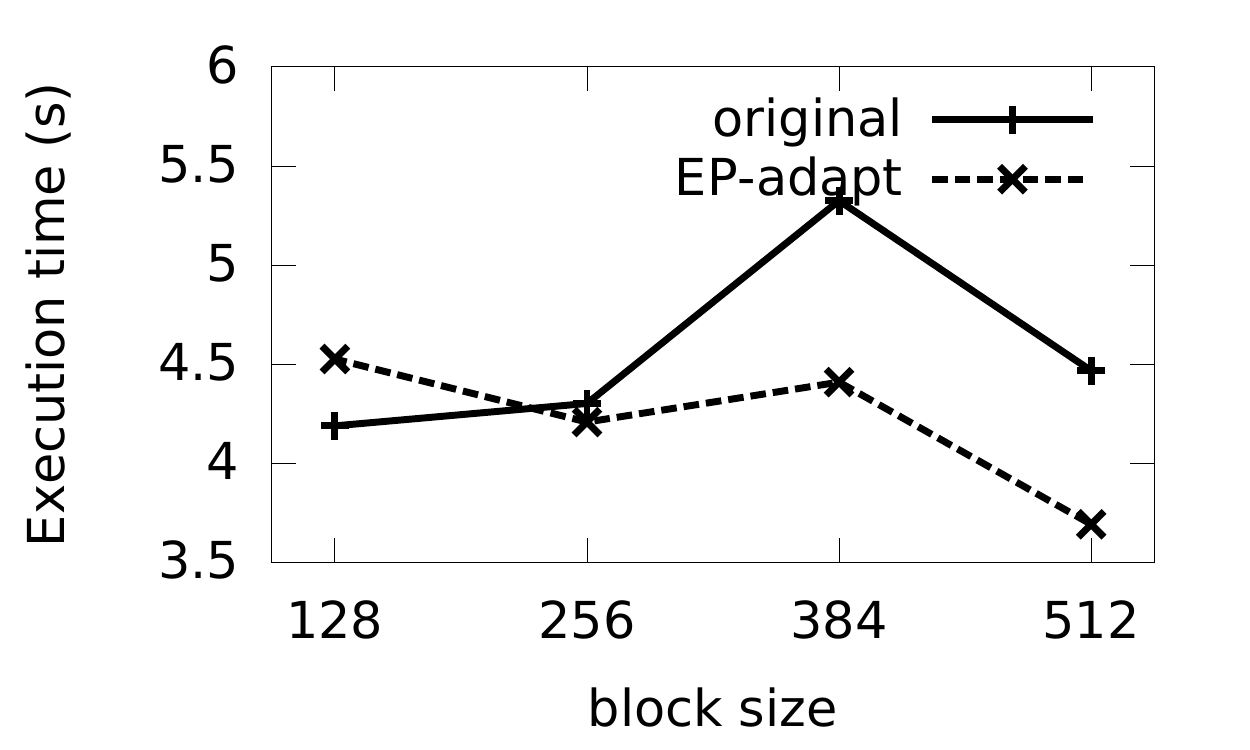}} \hfill
	\subfloat[cfd (m0.2M)]{\includegraphics[width=0.25\textwidth]{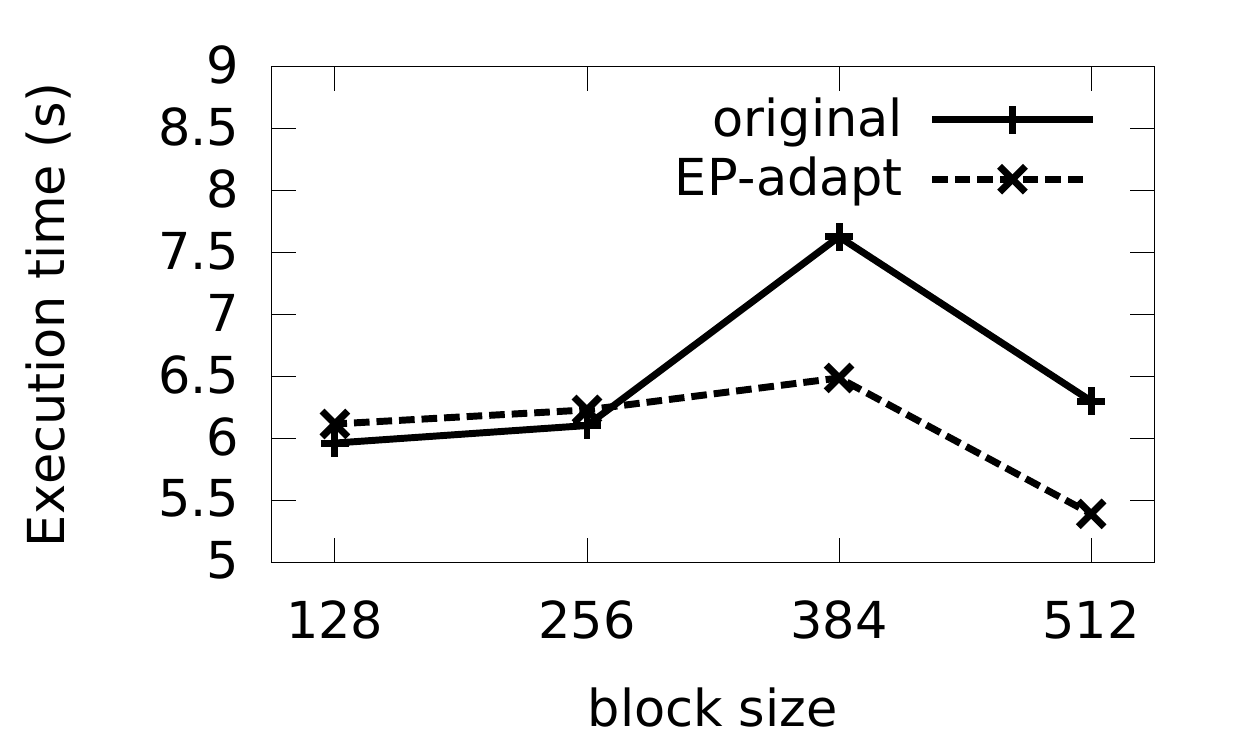}}
	\subfloat[gaussian]{\includegraphics[width=0.25\textwidth]{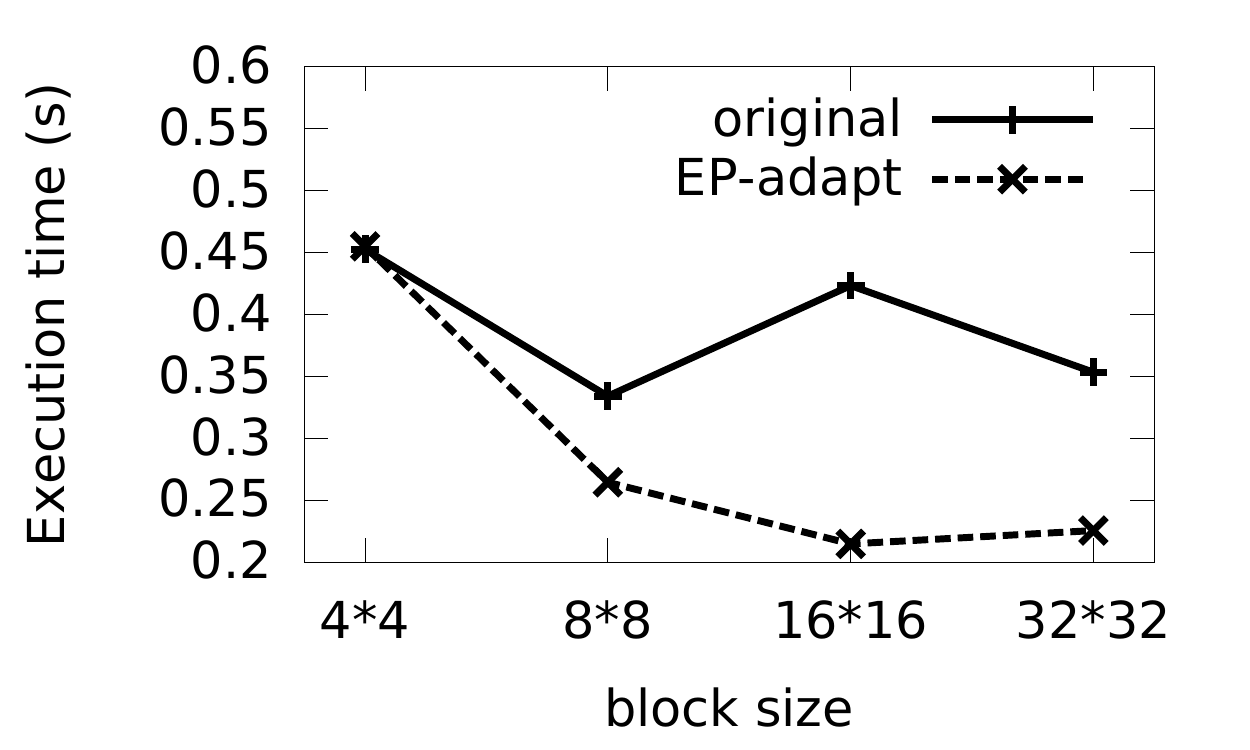}}
	\subfloat[particlefilter]{\includegraphics[width=0.25\textwidth]{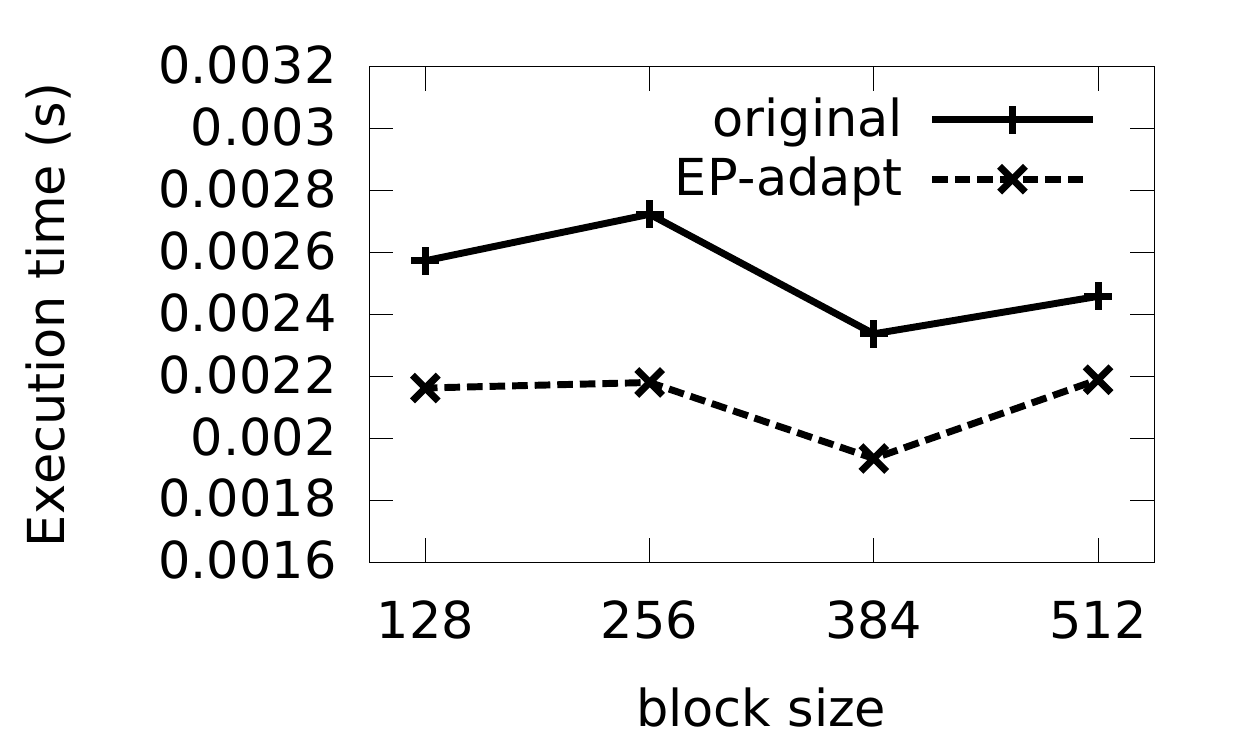}}
	\subfloat[streamcluster]{\includegraphics[width=0.25\textwidth]{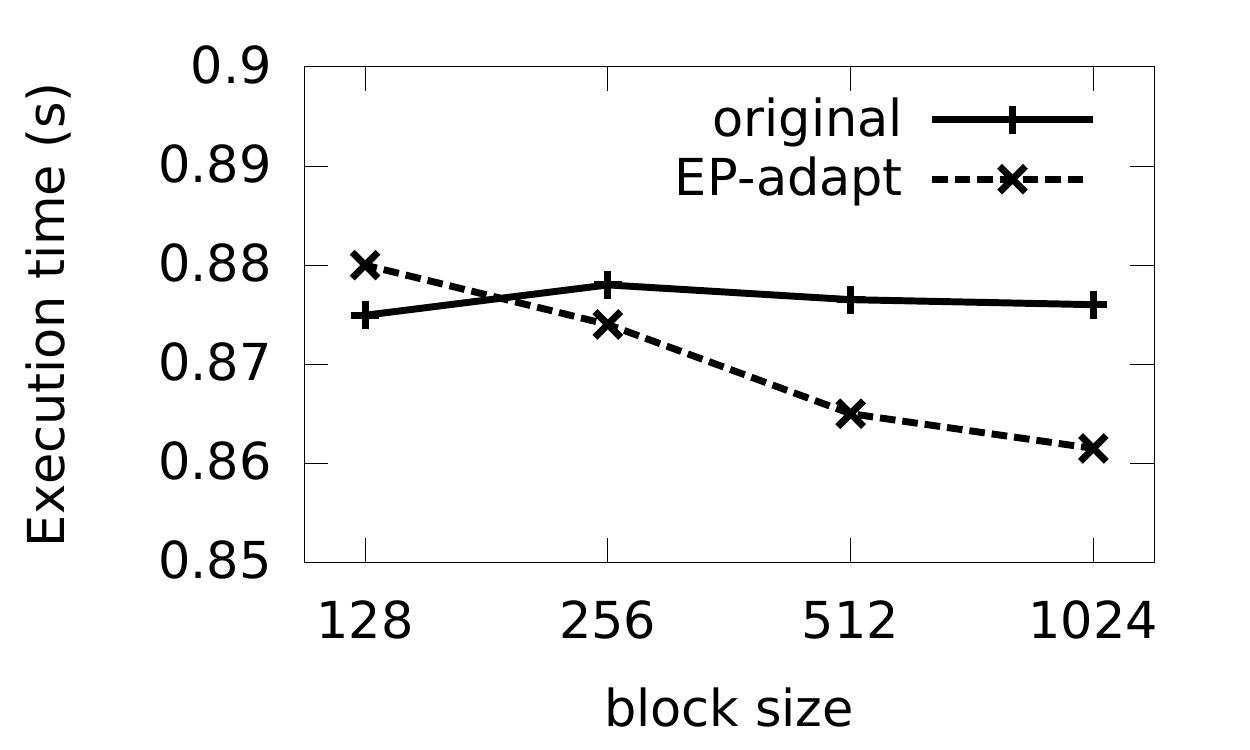}}
	\caption{Application performance on different block sizes.}
	\label{fig:appspec}
\end{figure*}

\subsection{General Workloads}
\label{subsec:app}

Figure \ref{fig:appspec} shows the performance of the six Rodinia applications under various thread block sizes. For every benchmark, we test performance using four thread block sizes, 128, 256, 384 and 512, unless the program does not allow such a thread block size.
For instance, \emph{gaussian} only allows square thread block sizes.
The original runtime is denoted as {\em original} in Figure \ref{fig:appspec}. The runtime of our EP model with overhead control is denoted as \emph{EP-adapt}.

\begin{figure}
	\centering
	\includegraphics[width=0.4\textwidth]{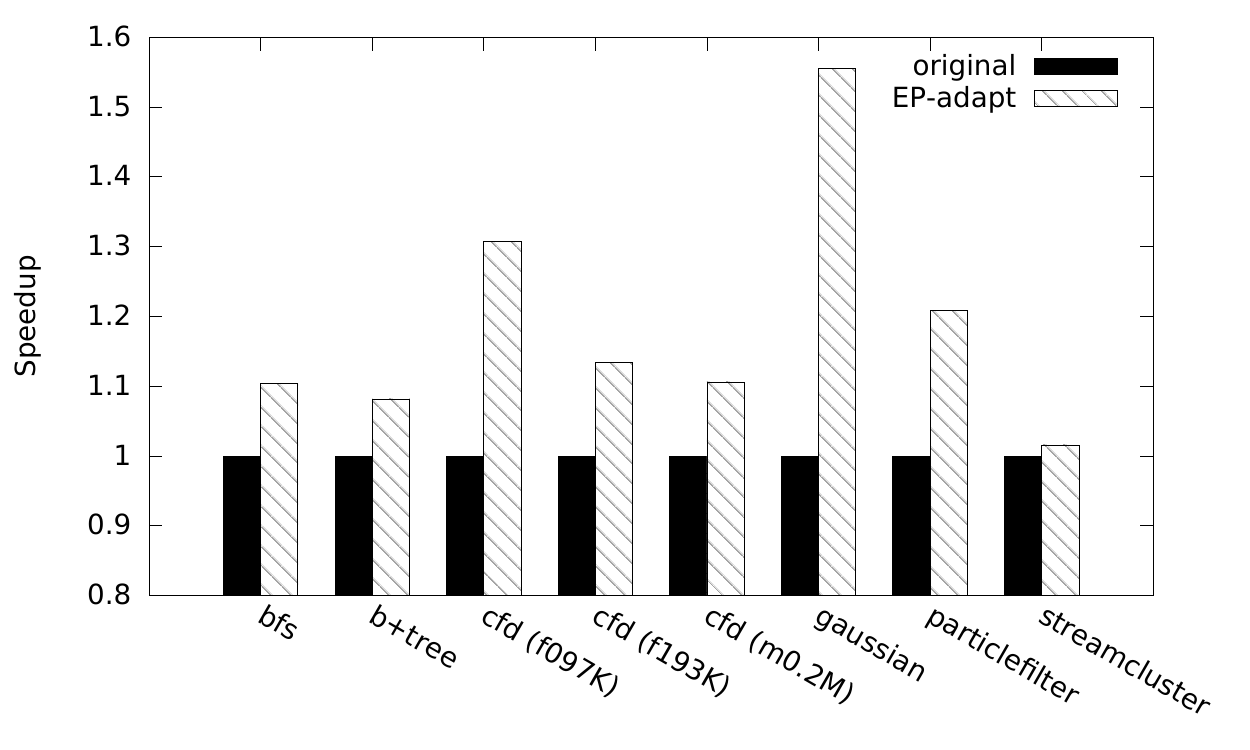}
	\caption{Application performance summary.}
	\label{fig:appsummary}
\end{figure}

First, we can see that in most cases our optimized version outperforms the original version when the thread block size is fixed. The maximum speedup is 1.97 times for {\em gaussian} at the thread block size 256.
For {\em streamcluster}, the maximum speedup 1.7\% is for thread block
size 1024, which is lower than that of other benchmarks.
This is because \emph{streamcluster} has less data sharing compared to other benchmarks at runtime.
In \emph{streamcluster}, every thread holds a unique data point, and a set of threads access the same center point to perform distance calculation, which makes the average degree of data-affinity graph to be $\le 2$ (average data reuse).

Enlarging thread block size typically would enlarge the amount of sharing since more threads imply more data sharing opportunities.
This is particularly true for {\em cfd}. The larger the thread block size is, the better performance it is for the EP model.
However, it is not the case for all benchmarks.
For instance, the best performance of EP-adapt is not achieved at the largest thread block size for \emph{particlefilter} and \emph{b+tree}.
This is because the GPU performance is determined by many factors besides data sharing.
Many of these factors change with the thread block size, such as occupancy, cache contention, etc.
For instance, \emph{cfd} suffers from performance degradation at a block size of 384 because 384 is non-integral power of 2, and thus it is impossible to achieve the best potential occupancy of 2048 threads.

In Figure \ref{fig:appsummary}, we show the comparison between the best EP-adapt version and the best original version in terms of performance across different thread block sizes.
The data is normalized to the runtime of the best original version.
Note that the execution time of \emph{EP-adapt} version includes the optimization overhead.
We have significant performance gains, or at least no performance degradation, for all benchmarks with adaptive overhead control.

\begin{figure}[t]
	\centering
	\includegraphics[width=0.8\linewidth]{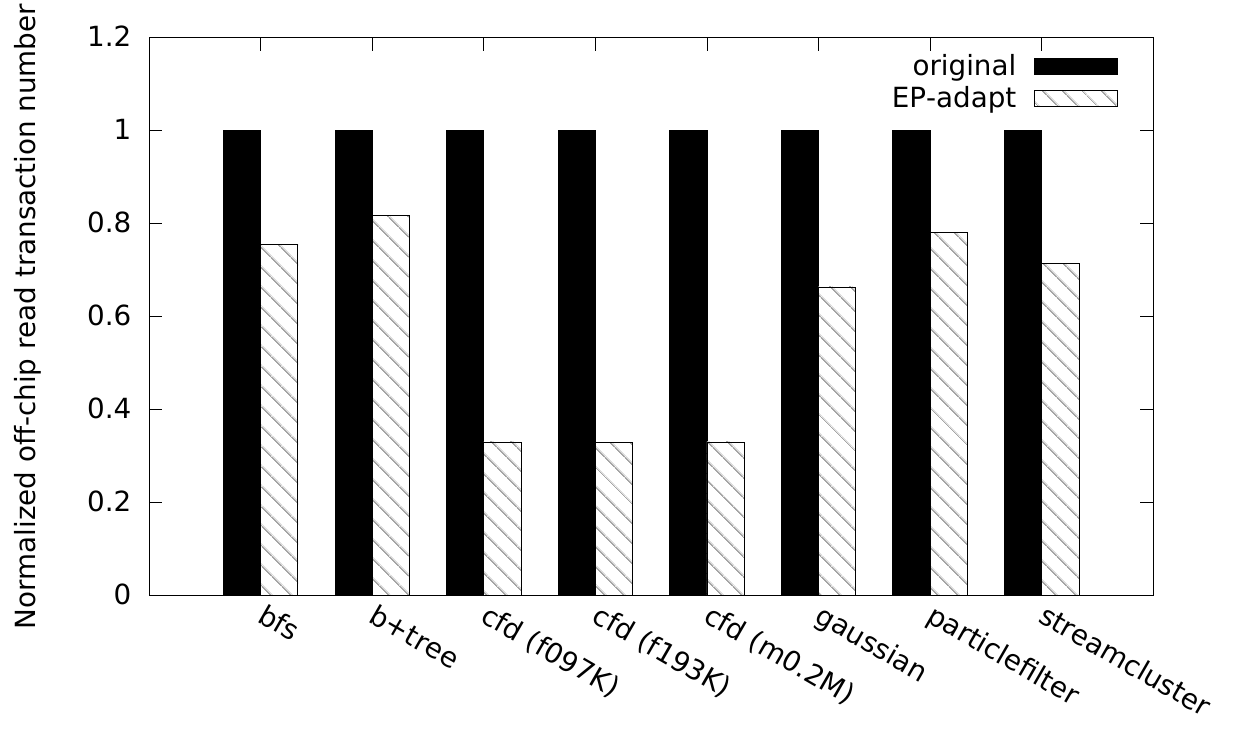}
	\caption{Read transaction number.}
	\label{fig:read_trans}
\end{figure}

Figure \ref{fig:read_trans} shows the normalized off-chip read transaction number measured by CUDA profiler for this set of benchmarks.
The reason why we do not show write transaction number is that there is no write sharing in these benchmarks.
All results are normalized to that of the {\em original} version at the same thread block size.
The results show that our technique can reduce off-chip memory transactions significantly, since more memory requests are satisfied by on-chip software or texture caches.


We also observed that using texture cache does not perform as well as software cache in most cases, which is similar to the observation for SPMV.

%% file: rel.tex
\section{Related Work}
\label{sec:rel}

\paragraph{CPU task partition for communication reduction}
Some models have been proposed to model data communication in traditional CPU parallel systems, and these models are used to reduce communication among processors through task scheduling.
Hendrickson et al. use vertex-partition based graph models where vertices represent tasks \cite{Hendrickson+:SIAMJSC00, Hendrickson+:PC00}.
The vertex-partition model cannot measure data communication accurately, and thus we use edges to represent tasks instead.
Hypergraph models \cite{karypis1998hmetis, catalyurek1999patoh} can model communication cost accurately.
The main drawback of the hypergraph model is its large overhead, as we demonstrate in Section \ref{sect:tech:compare}, which makes it infeasible for GPU computing.
Pregel \cite{malewicz2010pregel} and GraphLab \cite{low2012distributed} introduce two parallel computation models based on message passing and shared memory, respectively.
They always assign all computation of one vertex to one processor due to limitations of the programming model, while our model allows computation to be scheduled arbitrarily to achieve better partition quality and balance.
PowerGraph \cite{gonzalez2012powergraph} can distribute the computation of one vertex into several processors and proposes two edge partition methods for task scheduling.
One is to randomly assign edges, and the other, greedy based method assigns an edge to those partitions which already possess its endpoints.
These simple edge partition schemes produce low-quality partitions and are thus inapplicable for GPU task partitioning, as demonstrated in Section \ref{sect:tech:compare}.

Ding and Kennedy propose using runtime transformation for improving memory performance of irregular programs \cite{Ding+:PLDI99}, but it is heuristics based and there is no rigorous model for data reuse.
Bondhugula et al. introduce an automatic source-to-source transformation framework to optimize data locality, and they formulate data locality problem with polyhedral model \cite{Bondhugula+:PLDI08}, which only works for regular applications with affine memory indices.

\paragraph{GPU task partition for divergence elimination}
The dynamic task scheduling work for GPU memory performance optimization is mainly in memory coalescing rather than data sharing.
Zhang et al. propose to dynamically reorganize data and thread layout to minimize irregular memory accesses \cite{Zhang+:ASPLOS11}.
Wu et al. also propose two data reorganization algorithms to reduce irregular accesses \cite{Wu+:PPoPP13}.
However, these papers do not address the data sharing problem in GPU caches.

\paragraph{Other GPU software optimization techniques}
Many compiler techniques are proposed to achieve better utilization of GPU memory.
For affine loops, Baskaran et al. use a polyhedral compiler model to reduce non-coalesced memory accesses and bank conflicts in shared memory \cite{Baskaran+:ICS08}.
Jia et al. propose to characterize data locality and then guide GPU caching \cite{Jia+:ICS12}.
The limitation of these compiler methods is that they cannot address global data sharing among threads.

Some research uses hints provided by programmers to help compilers improve GPU memory performance.
CUDA-lite tunes shared memory allocation via annotations \cite{Ueng+:LCPC08}.
hiCUDA seeks to automate shared memory allocation with the help of programmer specified directives \cite{Han+:TPDS11}.

There are also application-specific studies for sparse matrix vector multiplication.
Bell and Garland discuss data structures of SPMV for various sparse matrix formats \cite{Bell+:NVIDIATR08}.
Choi et al. propose an automatic performance tuning framework for SPMV \cite{Choi+:PPoPP10}.
Volkov and Demmel analyze the bottleneck in dense linear algebra and optimize its performance by improving on-chip memory utilization and etc
\cite{Volkov+:SC08}.
But none of this handles the fundamental task scheduling problem for SPMV locality enhancement.
Nonetheless, these techniques are complementary to our technique. For instance, we can use auto-tuning to find out whether to use texture cache or software cache.